\documentclass[fleqn,usenatbib]{mnras}

\usepackage{newtxtext,newtxmath}

\usepackage[T1]{fontenc}

\DeclareRobustCommand{\VAN}[3]{#2}
\let\VANthebibliography\thebibliography
\def\thebibliography{\DeclareRobustCommand{\VAN}[3]{##3}\VANthebibliography}

\usepackage{graphicx}	
\usepackage{amsmath}	
\usepackage{longtable}  
\usepackage{caption}
\usepackage{fontawesome5}

\newcommand{\githublink}[2]{%
    \href{#1}{\textcolor{black}\faGithub\ #2}%
}

\title[Radius Evolution of 3I/ATLAS]{Assessment of the Mass Loss and Radius Change of 3I/ATLAS Based on Observed Production Rates}

\author[Frincke \& Seligman]{
Tessa T. Frincke,$^{1}$\thanks{E-mail: frincket@msu.edu}
Darryl Z. Seligman, $^{1}$
\\
$^{1}$Department of Physics and Astronomy, Michigan State University, East Lansing, MI 48824, USA\\
}

\date{Accepted XXX. Received YYY; in original from ZZZ}

\pubyear{2026}

\begin{document}
\label{firstpage}
\pagerange{\pageref{firstpage}--\pageref{lastpage}}
\maketitle

\begin{abstract}

Formed from the debris of planet formation, interstellar comets provide invaluable insights into the chemical compositions of planetary systems outside of our Solar System. Spectroscopic observations of 3I/ATLAS, the third interstellar object, reveal production of numerous volatiles and refractory species throughout its trajectory. In this paper we present a framework to calculate the change in radius of an object on an arbitrary trajectory at any point in its orbit, applicable to any small body experiencing mass loss. We next provide a comprehensive, machine readable table containing volatile and refractory production rates from all reported observations of 3I/ATLAS pre- and post-perihelion. Applying these equations to 3I/ATLAS, we calculate that it has lost $\sim$1.71--5.24 meters of its surface during its passage through the Solar System from gas alone, corresponding to $\sim$10$^{10}$ kg and $\sim$0.40--1.25\% of its initial mass. Considering dust-to-gas ratios from 1--4, we estimate surface erosion down to depths $>$10 meters. We present these calculations as approximations and lower limits based on discontinuous observational data with varying methodology and instrumentation. Conservative and optimistic estimates were calculated over a range of heliocentric distances defined by the onset of activity in reported observations and the typical onset of sublimation distance for each species, respectively. The reported production rates combined with the change in radius calculation can be used to estimate subsurface locations of various species within the nucleus of 3I/ATLAS.

\end{abstract}


\section{Introduction}

3I/ATLAS is the third macroscopic interstellar object to ever be detected in the inner Solar System \citep{Denneau2025}. After traversing the interstellar medium (ISM) over Gyr timescales \citep{Hopkins2025,Taylor2025a}, 3I/ATLAS was discovered on UT 2025 July 1 by the Asteroid Terrestrial-impact Last Alert System (ATLAS) survey at a heliocentric distance of $\sim$ 4.5 au \citep{Tonry2018a}. 3I/ATLAS adds to a growing population of interstellar objects including 1I/'Oumuamua and 2I/Borisov, discovered in 2017 and 2019, respectively \citep{Williams17, borisov_2I_cbet}.

Initial observations of 3I/ATLAS identified no significant light curve variation, a reddened spectral slope similar to D-type asteroids, and sustained faint activity that confirmed the cometary nature of the object \citep{Seligman2025a, Frincke2026, Puzia2025, Opitom2025, Kareta2025, Belyakov2025}. The first interstellar object, 1I/`Oumuamua, exhibited nongravitational acceleration \citep{Micheli2018} with no visible dust coma and photometric inactivity \citep{Ye2017, Fitzsimmons2018, Jewitt2017,Meech2017}. Conversely, 2I/Borisov, the second interstellar object, demonstrated typical cometary activity \citep{Jewitt2019b, Hui2020, Guzik2020, McKay2020, Fitzsimmons2019, Kim2020, Cremonese2020, Yang2021, Opitom2019, Kareta2019, Lin2020, Xing2020, Bannister2020, Bagnulo2021, Aravind2021, Deam2025, Cordiner2020, Bodewits2020}. Initial observations of 2I/Borisov reported physical properties broadly comparable to solar system comets \citep{Jewitt2019b, Hui2020, Guzik2020}. Post-perihelion observations of 2I/Borisov revealed a CO abundance relative to H$_2$O higher than typical solar system comets \citep{Bodewits2020, Cordiner2020}. 

Pre-discovery time-series photometry of 3I/ATLAS obtained from several facilities confirmed the onset of activity out to a heliocentric distance, $r_H$, of $\sim$ 6 au. \citet{Ye2025} reported observations obtained with the Zwicky Transient Facility (ZTF) starting on UT 2025 May 15 that confirmed 3I/ATLAS was active out to 6.5 au and maintained a persistent dust flow starting at $\sim$ 9 au. Photometry collected with the ATLAS network between UT 2025 March 29 and UT 2025 August 29 tracked the pre- and near-discovery brightness evolution of 3I/ATLAS \citep{Tonry2025}. Furthermore, independent analyses of pre-discovery photometry obtained with NASA's Transiting Exoplanet Survey Satellite (\textit{TESS}) between UT 2025 May and June confirmed 3I/ATLAS may have been active out to a $r_H$ of $\sim$ 6 au \citep{Feinstein2025, Martinez-Palomera2025}. Observations of 3I/ATLAS obtained with the NSF-DOE Vera C. Rubin Observatory between UT 2025 June 21 and the discovery date of UT 2025 July 1 also confirmed activity at large heliocentric distances \citep{Chandler2025}.

Observations of the inbound trajectory of 3I/ATLAS traced the onset of volatile and refractory emission in the coma. For example, \citet{Alvarez-Candal2025} estimated the first pre-perihelion upper limit of H$_2$O production from a detection of OH at 4.42 au; beyond where water ice typically sublimates at $<$ 3 au. H$_2$O ice was detected at $\sim$ 4 au from observations obtained by \citet{Yang2025} with the NASA Infrared Telescope Facility (IRTF). Furthermore, \citet{Xing2025} constrained water production rates from UT 2025 July 31 -- August 1 and UT 2025 August 18 -- 20 at 3.51 au from data obtained with the Ultraviolet/Optical Telescope (UVOT) instrument on the \textit{Neils Gehrels Swift Observatory}. \citet{Cordiner2025} reported detections of water ice and gas emission with the James Webb Space Telescope (\textit{JWST}) Near Infrared Spectrograph (NIRSPec) on UT 2025 August 6 at 3.32 au. \citet{Lisse2025b} also detected H$_{2}$O gas features at 3.26 au with the Spectro-Photometer for the History of the Universe, Epoch of Reionization, and Ices Explorer \textit{SPHEREx} spacecraft. In addition to water ice and gas, \citet{Cordiner2025} found the coma was enriched in CO$_2$ compared to CO and H$_2$O based on an unusually high CO$_2$/H$_2$O mixing ratio compared to solar system comets at the same $r_H$ of 3.32 au. \citet{Lisse2025a} similarly detected a CO$_{2}$ dominated and CO depleted coma from spectrophotometry of 3I/ATLAS obtained with \textit{SPHEREx} on UT 2025 August 1--15. For further discussion of this behavior in solar system comets see Section~\ref{sec:05}. \citet{Hinkle2025} estimated an upper limit on a production rate of CO from observations of 3I/ATLAS obtained between UT 2025 July 16--21 with the James Clerk Maxwell Telescope (JCMT). \citet{Li2026} further contributed to pre-perihelion volatile production rates with detections of OH, CO, HCN, and H$_2$O derived from radio observations obtained with the 13.7-meter Millimeter-Wave Telescope from UT 2025 August to September.

Pre-perihelion CN was detected in the coma of 3I/ATLAS at $\sim$ 4.47 au on UT 2025 July 2--5 contemporaneously by \citet{delaFuenteMarcos2025} and \citet{Alvarez-Candal2025}. Subsequent pre-perihelion observations reported detections of CN emission through UT 2025 September 25 at 1.85 au \citep{Hoogendam2025a, Rahatgaonkar2025, Salazar-Manzano2025, Hutsemekers2026a, Hoogendam2025b, Medler2026, Lisse2026,Paek2026}. Observations obtained on UT 2025 August 8--17 with the MDM Observatory measured a C$_2$-to-CN ratio that was indicative of strong carbon-chain depletion \citep{Salazar-Manzano2025}. Pre-perihelion observations of CN emission demonstrated an overall increase in production rates with decreasing heliocentric distance. \citet{Coulson2025} reported production rates for hydrogen cyanide (HCN) from observations conducted on UT 2025 August 7 and UT 2025 September 14 with the JCMT. \citet{Hinkle2025} derived a steeper slope for the production of HCN than is typical for solar system comets between 4.01--3.97 au and estimated an upper limit production rate. \citet{Roth2025} found asymmetrical production rates with depleted HCN and enhanced CH$_3$OH in the sunward region of the coma from data collected with the Atacama Large Millimeter/Sub-millimeter Array (ALMA) from UT 2025 August 28 to UT 2025 October 1.

Analyses of post-perihelion production rates of volatiles and refractory species in the coma of 3I/ATLAS reveal perihelion asymmetries that may be indicative of structural inhomogeneities such as volatile reservoirs in sub-surface layers. For example, \citet{Zhao2026} found the overall trend in the post-perihelion production rates derived for CN, Ni, Fe and H$_2$O show shallower declining slopes compared to pre-perihelion detections. From the same observations obtained with the Xinglong 2.16-meter and Li-jiang 2.4-meter Telescopes between UT 2025 December 2 and UT 2026 January 20, \citet{Zhao2026} measured production rates that revealed a more enriched C$_2$ post-perihelion coma. Similarly, \citet{Medler2026} measured an order of magnitude increase in CN production from observations spanning from UT 2025 July to November. Despite asymmetries in the overall production rate trend with heliocentric distance, \citet{Lisse2026} estimated an H$_2$O gas production rate that was $\sim$40 times larger than pre-perihelion observations, implying a reservoir of water ice was exposed after perihelion passage. Post-perihelion abundances reveal a change from H$_2$O to CO production as the primary driver of activity in the coma of 3I/ATLAS approaching 3 au \citep{Biver2026}. To illustrate, \citet{Shinnaka2026} estimated that the abundance ratio of CO$_2$/H$_2$O decreased since pre-perihelion measurements. Moreover, \textit{SPHEREx} observations taken at $r_H =$ 2--2.2 au found a higher Q(CO)/Q(H$_2$O) ratio of 0.55 compared to a Q(CO$_2$)/Q(H$_2$0) ratio of 0.23 indicative of a CO-dominated post-perihelion coma \citep{Lisse2026}. This behavior was confirmed with spatial-spectral mapping from observations obtained on UT 2025 December 22-23 with the \textit{JWST} NIRSpec Integral Field Unit (IFU) that indicated that CO was the most abundant species \citep{Roth2026}.

Further detections of volatiles and refractory species trace the post-perihelion activity of 3I/ATLAS. \citet{Combi2025} derived a water production rate on UT 2025 December 9 that declined by an order of magnitude compared to a near-perihelion measurement they obtained on UT 2025 November 6 with the SOHO/SWAN instrument. Furthermore, observations with the Solar Wind ANisotropies (SWAN) camera on the Solar and Heliosphere Observatory (SOHO) of 3I/ATLAS from UT 2025 November and December confirmed a post-perihelion asymmetry in water production rates \citep{Tan2026}. \citet{Hutsemekers2026b} tracked the post-perihelion Ni and Fe emission from 3I/ATLAS using the UVES instrument on the VLT spanning from UT 2025 December 4 to UT 2026 February 22 and reported a more gradual decrease in production rates. \citet{Hoogendam2026} reported production rates of Ni, Fe, CN, C$_2$, and C$_3$ derived from IFU spectroscopic measurements of 3I/ATLAS taken with the Keck Cosmic Web Imager (KCWI) on UT 2025 November 16. \citet{Belyakov2026} revealed the first detection of CH$_4$ and a reduction in overall outgassing from spectroscopic measurements of 3I/ATLAS obtained with \textit{JWST} Mid-Infrared Instrument (MIRI) on UT 2025 December 15, 16, and 27. Continued monitoring and analyses of the activity of 3I/ATLAS have been reported contemporaneously with those outlined here \citep{Furuya2026, Ferellec2026, Cochran2026, SantanaRos2026}.

Interstellar objects in general, and 3I/ATLAS specifically, offer the opportunity to potentially measure the chemical composition of primordial planetary building blocks. However, 3I/ATLAS may have undergone extreme processing by galactic cosmic rays over $\sim$ Gyrs in the ISM before being detected in the inner Solar System, which would have altered its composition, similar to mechanisms invoked to explain 1I/`Oumuamua's acceleration \citep{Bergner2023}. \citet{Maggiolo2025} posited that 3I/ATLAS was processed via radiolytic conversion of CO to CO$_2$ by galactic cosmic rays (GCR) in the ISM down to a depth of $\sim$ 15--20 m \citep[see also \ ][]{Gronoff2020, Maggiolo2020}. The depletion of CO in the coma of dynamically new long period solar system comets due to GCR processing imply specific formation pathways of 3I/ATLAS that may have lead to the observed CO$_2$ dominated pre-perihelion activity \citep{Harrington-Pinto2022, AHearn2012}. For a more in-depth discussion of the GCR processing of the surface layers of comets see Section~\ref{sec:05}. The derived kinematic age of $\sim$3--11 Gyr \citep{Taylor2025a} and inferred isotopic age of 11--12 Gyr \citep{Cordiner2026} of 3I/ATLAS emphasizes the need to address the potential for surface processing in the ISM after formation. Therefore, estimates of surface erosion are vital to determine the observability of minimally processed materials in the coma from sublimation driven activity of 3I/ATLAS outlined above.

The paper is organized as follows: in Section~\ref{sec:02}, we outline a methodology for estimating the change in nuclear radius of any small body for an arbitrary orbit and mass loss rate. In Section~\ref{sec:03}, we provide a summary of production rates of volatiles and refractory species from all reported pre- and post-perihelion measurements. In Section~\ref{sec:04}, we estimate the change in radius of 3I/ATLAS over its trajectory in the Solar System and provide a table of mass loss derived over conservative and optimistic ranges in position and the contributions of certain volatile species. We further assess the change in radius in consideration of a range of dust-to-gas ratios (D/G) applied to total mass loss rates. These estimates are based on observations from various instruments with differing observational techniques and only account for the gaseous species detected in the observations. In Section~\ref{sec:05}, we discuss the results and summarize our conclusions in Section~\ref{sec:06}.

\section{Change in Radius for Arbitrary Orbit and Mass Loss Function}
\label{sec:02}

In this section we present a methodology to estimate the change in radius of a small body experiencing mass-loss at an arbitrary location of an arbitrary orbit in the Solar System. This methodology is useful for quantifying the erosion of processed layers from the surface of 3I/ATLAS. This calculation assumes a spherically symmetric geometry for the nucleus, but could be updated to arbitrary shapes in future work.

\subsection{Change in Radius Formulation}
\label{subsec:2.1}

We assume that the mass loss rate is directly proportional to the rate of change of the volume of the nucleus. The mass loss rate, $\dot{m}$, for the generic small body is then:
\begin{equation}
    \dot{m} = \rho\frac{dV}{dt}\,.
    \label{eq:01}
\end{equation}
In Equation~(\ref{eq:01}), $\rho$ is the bulk density of the nucleus. Assuming a spherical geometry for the nucleus, the volume, $V$, and the rate of change of the volume is given by,
\begin{equation}
    \frac{dV}{dt} = 4\pi R^{2}\frac{dR}{dt}\,. 
    \label{eq:02}
\end{equation}
In Equation~(\ref{eq:02}), $R$ is the radius of the nucleus. Substituting Equation~(\ref{eq:02}) into Equation~(\ref{eq:01}) gives the rate of change of the nuclear radius during its passage in the Solar System as,
\begin{equation}
        \frac{dR}{dt} = {\dot{m}}\bigg/{4 \pi R^{2}\rho}\,.
        \label{eq:03}
\end{equation}
After separation of variables and integration over an arbitrary range of time, $\Delta\tau$, we solve for the reduction of radius with time given as,
\begin{equation}
    R_f = \bigg(R_0^{3} + \big(3\dot{m}\Delta\tau\big)\bigg/\big({4\pi\rho} \big)\bigg)^{\frac{1}{3}} \,.
    \label{eq:04}
\end{equation}
In Equation~(\ref{eq:04}), $R$$_0$ is the initial nuclear radius at some initial time, $\tau$$_0$, and $R$$_f$ is the radius after some time, $\Delta\tau$. Here we assume that $\dot{m}$ remains constant for simplicity; later we will introduce more complex mass loss functions to the derivation that can be applied more generally.

\begin{figure}
    \centering
    \includegraphics[width=1.\linewidth]{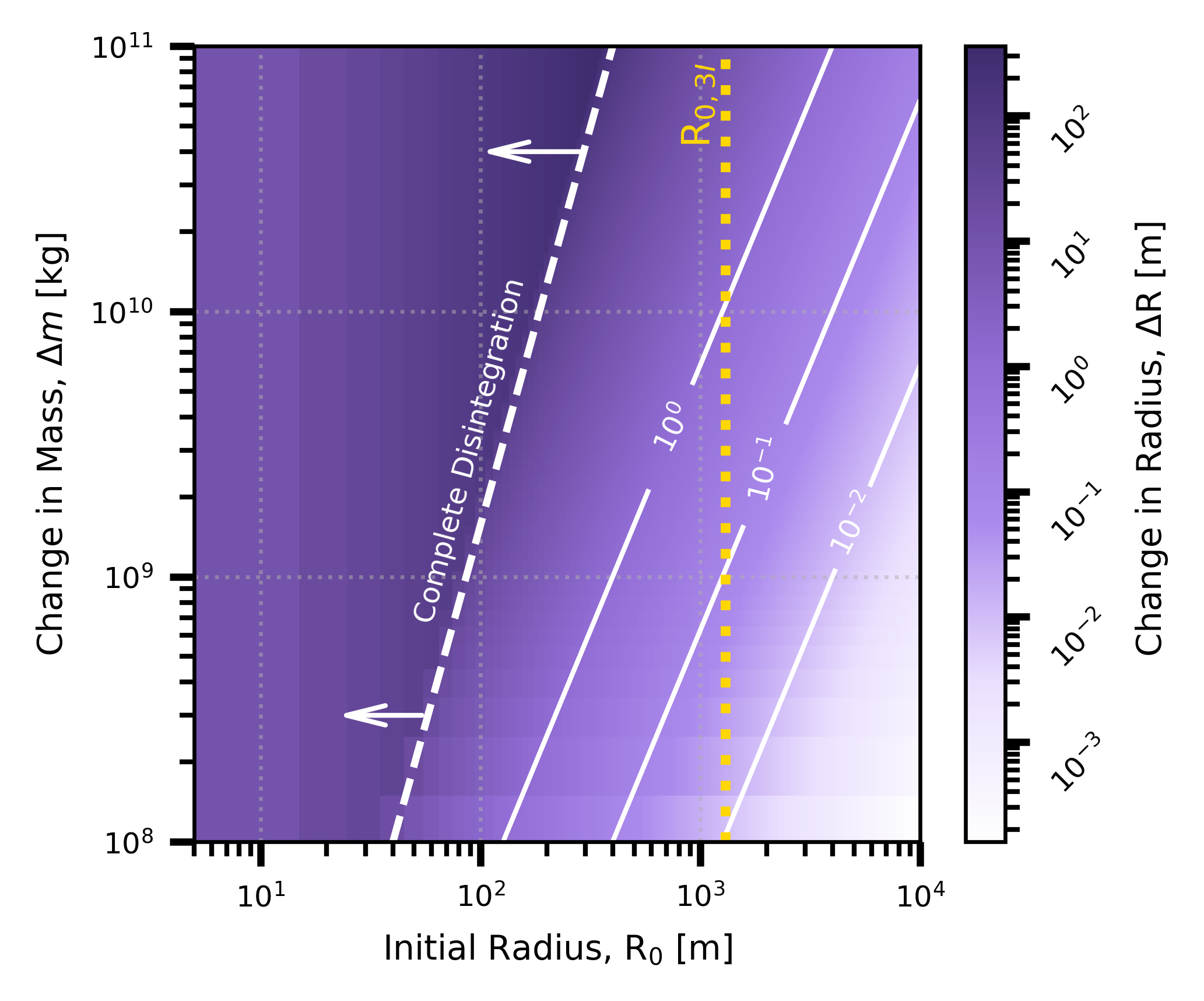}
    \caption{The change in radius of a small body sensitively depends on its initial radius. Small bodies with larger initial radii require significant mass loss to have appreciable nuclear erosion. Contour lines indicate constant change in radius given a constant total change in mass,  $\Delta m$, initial radius, $R$$_0$, and assumed density of 0.5 g cm$^{-3}$. The area left of the dashed line indicates the regime where the object loses all of its mass. The vertical dotted line indicates the nuclear radius of 3I/ATLAS derived by \citet{Hui2026}.}
    \label{fig:dM_Ro_DRcolorbar}
\end{figure}

Small bodies with smaller initial radii exhibit significant radius loss. Moreover, the overall change in radius is extremely sensitive to the initial radius of the object. This sensitivity is visualized in Figure~\ref{fig:dM_Ro_DRcolorbar} where the change in radius, $\Delta R$, is computed using Equation~\ref{eq:04} over a range of total change in mass, $\Delta m$, values and initial comet radius, $R_0$. The $\Delta R$ in Figure~\ref{fig:dM_Ro_DRcolorbar} was computed at an arbitrary time to emphasize the sensitivity of the calculation on $R_0$. We assume a typical comet density of 0.5 g cm$^{-3}$ based on average densities from observations of solar system comets \citep{AHearn2011, Sierks2015}. The dashed line separates the regime of mass loss and initial radii over which the change in radius is equal to the initial radius, indicative of complete disintegration of the nucleus.

\subsection{Change in Radius with Constant Mass}
\label{subsec:2.2}

Here we calculate the change in radius at an arbitrary position, $r$, in a small bodies' orbit. To achieve this, we calculate a solution, similar to Equation~(\ref{eq:04}), for the nuclear radius at a given true anomaly for a small body in either a circular, eccentric, or hyperbolic orbit. The velocity vector, $\dot{\mathbf{r}}$, and angular momentum vector, $\mathbf{h}$, assuming a two-body system are given as,
\begin{equation}
\dot{\mathbf{r}} = \dot{r} \mathbf{\hat{r}} + r\dot{\theta} \boldsymbol{\hat{\theta}} \,,
\label{eq:05} 
\end{equation}
and
\begin{equation}
    \mathbf{h} = \mathbf{r} \times \dot{\mathbf{r}} = r^2 \dot{\theta} \,.
    \label{eq:06} 
\end{equation}
In Equation~(\ref{eq:06}), $\theta$ is the true anomaly. The position of the small body in its orbit is determined by:
\begin{equation}
    r(\theta) = \frac{p}{1+e\cos{\theta}} \,.
    \label{eq:07} 
\end{equation}
In Equation~(\ref{eq:07}), $p$ is the semilatus rectum and $e$ is the eccentricity of the object. Rearranging Equation~(\ref{eq:06}) gives $\dot{\theta} = h/r^2$, where the magnitude of the angular momentum vector is defined as $h = \sqrt{p\mu}$ and $\mu$ is the equation of relative motion between two bodies. The differential equation relating change in radius to the change in true anomaly is given by,
\begin{equation}
     \frac{dR}{d\theta} = \frac{\dot{m}}{4\pi\rho}\frac{r^2}{h} \,.
     \label{eq:08} 
\end{equation}
Therefore, the radius change can generically be calculated using:
\begin{equation}
    \int_{R_0}^{R_f} R^2 \,dR  = \int_{\theta_0}^{\theta_f} \bigg( \frac{\dot{m}}{4\pi\rho}\frac{r^2}{h}    \bigg)\, d \theta \,.
    \label{eq:09} 
\end{equation}
In Equation~(\ref{eq:09}) we assume that $\theta_0$ and $\theta_f$ span some range of true anomaly over which the final nuclear radius, $R_f$, will be evaluated.

 \subsubsection{Circular Orbit}
 \label{subsubsec:2.2.1}
 
For the circular orbit with $e = 0$, Equation~(\ref{eq:07}) reduces to $r = a$, where $a$ is the semi-major axis of the orbit. Integrating Equation~(\ref{eq:09}) assuming $\theta_0 = 0$ at perihelion, gives the solution for the radius for a circular orbit with a generic constant mass loss, $\dot{m}$, as,
\begin{equation}
     R_f = \bigg[R_0^{3} + \dot{m}\bigg(\frac{3 a^2}{4\pi \rho h}\bigg) \theta\bigg]^{\frac{1}{3}} \,.
    \label{eq:10} 
\end{equation}

 \subsubsection{Eccentric Orbit}
 \label{subsubsec:2.2.2}

For the eccentric orbit case where $0<e<1$, Equation~(\ref{eq:07}) does not reduce and is substituted directly into Equation~(\ref{eq:09}) with $p = a(1-e^2)$,
\begin{equation}
    \int_{R_0}^{R_f} R^2 \,dR  = \int_{0}^{\theta} \bigg(\frac{h^3}{\mu^2}\frac{\dot{m}}{4\pi \rho}\bigg)\frac{d \theta}{{(1+e\cos{\theta})^2}} \,.
    \label{eq:11} 
    \end{equation}
Equation~(\ref{eq:11}) gives the solution for the final radius integrated over a given range of true anomaly from perihelion passage as,
\begin{equation}
\begin{split}
     R_f = \bigg[R_0^{3} -  \bigg(\frac{h^3}{\mu^2}\frac{3\dot{m}}{4\pi \rho}\bigg) \frac{1}{(1-e^2)^{\frac{3}{2}}} \bigg. \\
     \bigg. \times \bigg(2\arctan{\sqrt{\frac{1-e}{1+e}}} \tan{\big(\frac{\theta}{2}\big)} - \frac{e\sqrt{1-e^2} \sin{\theta}}{1 - e\cos{\theta}}\bigg) \bigg]^{\frac{1}{3}} \,.
\end{split} 
\label{eq:12}
\end{equation}

\subsubsection{Hyperbolic Orbit}
\label{subsubsec:2.2.3}

Similarly, for a hyperbolic trajectory where $e>1$, Equation~(\ref{eq:11}) is instead substituted with $p = a(e^2-1)$. The angle between incoming and outgoing asymptotes of the hyperbolic trajectory as $r$ approaches infinity, $\theta_{min}$ and $\theta_{max}$, define the bounds of integration when evaluating the entire hyperbolic trajectory of the small body in the Solar System. The maximum angle, $\theta_{max}$, is derived from the limits of Equation~(\ref{eq:07}) and is defined as,
\begin{equation}
    \theta_{max} = 2\cos^{-1}{\left(-\frac{1}{e}\right)} \,. 
    \label{eq:13} 
\end{equation}
The solution to integration from perihelion to a given true anomaly gives the final radius through a given range of true anomaly for a small body on a hyperbolic trajectory:
\begin{equation}
\begin{split}
     R_f = \Bigg[R_0^{3} - \bigg(\frac{h^3}{\mu^2}\frac{3 \dot{m}}{4\pi \rho}\bigg) \frac{1}{e^2-1} \Bigg[\frac{e\sin{\theta}}{1+e\cos{\theta}} - \frac{1}{\sqrt{e^2-1}} \bigg. \bigg. \\ \bigg. \bigg. \times \ln \bigg(\frac{\sqrt{e+1}+\sqrt{e-1}\tan{\frac{\theta}{2}}}{\sqrt{e+1}-\sqrt{e-1}\tan{\frac{\theta}{2}}}\bigg)  \Bigg]\Bigg]^{\frac{1}{3}} \,.
\end{split} 
\label{eq:14}
\end{equation}

\begin{figure}
    \centering
    \includegraphics[width=1.\linewidth]{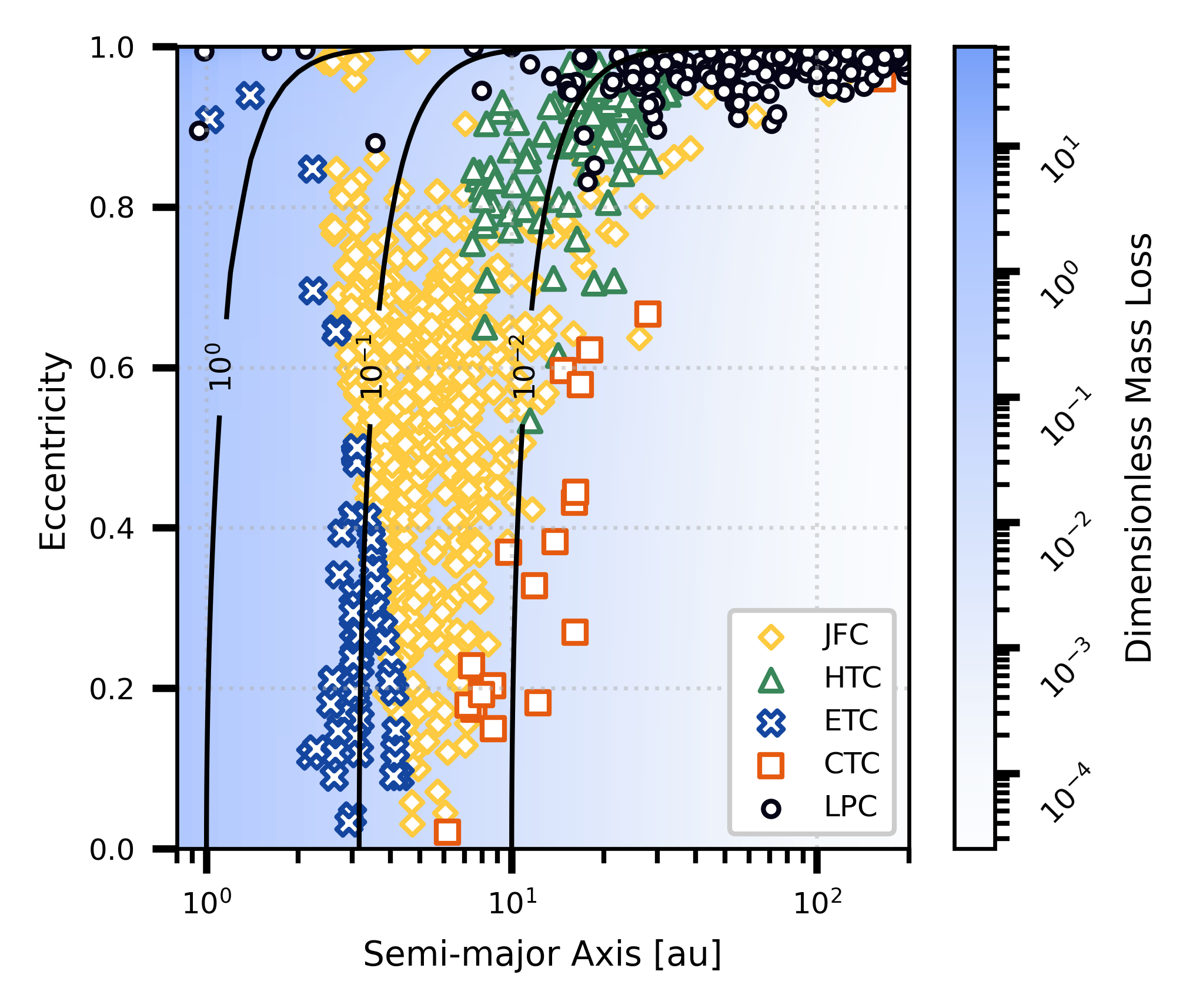}
    \caption{Solar System comets with larger eccentricities and smaller semi-major axes exhibit more mass loss over a single orbit. Contours are lines of constant orbit-averaged dimensionless mass loss. Jupiter-Family Comets (JFC) are represented as diamonds, Halley-Type Comets (HTC) as triangles, Encke-Type Comets (ETC) as crosses, and Chiron-Type Comets (CTC) as squares, and Long-Period Comets (LPC) or other as circles.}
    \label{fig:placeholder_e_a}
\end{figure}

Solar System comets with larger eccentricities and smaller semi-major axes experience more relative mass loss on average over a single orbit. In Figure~\ref{fig:placeholder_e_a} we show the orbits of solar system comets queried from JPL Small-Body Database \citep{giorgini2015nasa} plotted with contours as lines of constant orbit-averaged dimensionless mass loss proportional to $r^{-2}$.

\subsection{Change in Radius with Inverse Square Law Mass Loss Function}
\label{subsec:2.3}

Here we present solutions for the erosion of the surface of small body on a circular, eccentric, or hyperbolic orbit assuming an inverse square mass loss given as,
\begin{equation}
    \dot{m} = A\ \big(r^{-2}\big) \,. 
    \label{eq:15}
\end{equation}
In Equation~(\ref{eq:15}), $A$ is a constant moss loss coefficient. The production rate, and in turn, the mass loss rate from the outgassing of volatiles can be approximated by an inverse-square law with heliocentric distance as $r_H^{-2}$. We refer to heliocentric distance as it pertains to the Solar System, r$_H$, interchangeably with a generic distance, r, as defined in Equation~(\ref{eq:07}). The validity of this approximation is dependent on the molecular species and its sublimation temperature. For example, water will have a steeper inverse power law dependency on heliocentric distance than will CO and CO$_2$ over a range of $\sim$ 1--4 au due to higher sublimation temperatures \citep{Cowan1979, Marshall2019}. Mass
lost due to the acceleration experienced by dust also has a known
inverse square law dependence \citep{Jewitt2025-nongrav}. For use of the following solutions in Section~\ref{subsec:2.3}, the contributing volatile species should be considered to determine whether the $r_H^{-2}$ approximation defined in Equation~(\ref{eq:15}) is a reasonable mass loss function for the intended use-case.

\subsubsection{Circular Orbit}
\label{subsubsec:2.3.1}

For a circular orbit, the mass loss of a nucleus is defined as $\dot{m}=A/a^2$, and substituting into Equation~(\ref{eq:10}) for $\dot{m}$ gives,
\begin{equation}
     R_f = \bigg[R_0^{3} - \bigg(\frac{3A}{4\pi \rho h}\bigg) \theta \bigg]^{\frac{1}{3}} \,.
     \label{eq:16}
\end{equation}

\subsubsection{Eccentric Orbit}
\label{subsubsec:2.3.2}

For an eccentric orbit, the final radius is similarly calculated using a mass loss rate proportional to r$^{-2}$,
\begin{equation}
    \dot{m} =A \,\bigg(\,\frac{(1+e\cos{\theta)^2}}{a^2(1-e^2)^2}\,\bigg)\,.
    \label{eq:17}
\end{equation}
The final radius can be calculated using,
\begin{equation}
     R_f = \left[R_0^{3} - \left(\frac{3A}{4\pi \rho h_{ecc}}\right) \theta \right]^{\frac{1}{3}} \,.
     \label{eq:18}
\end{equation}
In Equation~(\ref{eq:18}) the magnitude of angular momentum for an eccentric orbit, $h_{ecc}$, is defined as:
\begin{equation}
    h_{ecc} = \sqrt{a(1-e^2)\mu}\, .
    \label{eq:19}
\end{equation}

\subsubsection{Hyperbolic Orbit}
\label{subsubsec:2.3.3}

And finally for a hyperbolic trajectory the inverse square mass lass function is,
\begin{equation}
    \dot{m} = A \,\bigg(\,\frac{(1+e\cos{\theta)^2}}{a^2(e^2-1)^2}\,\bigg)\,.
    \label{eq:20}
\end{equation}

The final radius for a hyperbolic trajectory is given as the same solution as for an eccentric orbit in Equation~(\ref{eq:18}), but $h$ is instead defined as, 
\begin{equation}
    h_{hyp} = \sqrt{a(e^2-1)\mu}\, .
    \label{eq:21}
\end{equation}

\begin{table*}
    \centering
    \caption{Power law indices, $\beta$, relating published production rates for 3I/ATLAS reported in \ref{tab:Q-table} to position or heliocentric distance, $r_H$, for pre- and post-perihelion. Columns $m_{ \textrm{pre, con}}$ and $m_{ \textrm{pre, opt}}$ are the conservative and optimistic estimates for total mass loss, respectively, from pre-perihelion observations. Similarly for post-perihelion observations $m_{ \textrm{post, con}}$ and $m_{ \textrm{post, opt}}$ are the conservative and optimistic estimates for total mass loss.  Entries with dashes indicate insufficient data to determine a fit. All total mass loss values $m$ are listed as a factor of $10^6$ kg.}
    \begin{tabular}{ccccccccccc}
        \hline\hline
Volatile      & $\beta_{\textrm{pre}}$  & $m_{ \textrm{pre, con}}$  & $r_{H, \textrm{ pre, con}}$&$m_{\textrm{pre, opt}}$    &$r_{H, \textrm{ pre, opt}}$& $\beta_{\textrm{post}}$  &$m_{\textrm{post, con}}$&r$_{H,\textrm{ post,con}}$&$m_{\textrm{post, opt}}$    &$r_{H, \textrm{ post, opt}}$\\
              &                & 10$^6$ [kg]          &[au]&10$^6$ [kg]  &[au]          &                 &10$^6$ [kg]   &[au]&10$^6$ [kg]  &[au]\\
\hline
H$_2$O& (8.713 $\pm$ 1.681)&813.8&3.51-1.96&33607&3.26--1.36& (8.253 $\pm$ 0.560)&13734&1.36--2.54&13910&1.36--3.26\\
CO$_2$& ---&---&---&---&---& (1.504 $\pm$  0.826)&1783&1.36--2.54&3370&1.36--12.25\\
CO& (3.755 $\pm$ 2.821)&231.4&3.90-2.01&1647&23.3--1.36& (1.033 $\pm$ 0.604)&1251&1.36--2.40&4083&1.36--23.30\\
 CN& (5.250 $\pm$ 0.485)&3.842&4.41--1.85&28.45&5.02--1.36& (4.619 $\pm$ 0.706)&7.818&1.50--3.29&17.53&1.36--5.02\\
Ni            & (6.167 $\pm$ 0.997)&0.2370&3.78--1.85&2.685&11.7--1.36& (5.452 $\pm$ 1.027)&2.778&1.50--3.85&7.383&1.36--11.7\\
Fe            & (10.09 $\pm$ 0.791)&0.0595&2.64--1.85&2.656&6.72--1.36& (7.462 $\pm$  0.838)&1.162&1.50--3.50&3.889&1.36--6.72\\
C$_2$         & (4.709 $\pm$ 1.066)&0.9987&3.16--1.85&6.915&8.68--1.36& (8.897 $\pm$ 0.480)&0.2324&1.50--2.42&8.331&1.36--8.86\\
 C$_3$& ---& ---& ---& ---& ---& (4.382 $\pm$ 1.437)& 0.4948& 1.50--2.42& 1.273&1.36--10.85\\
 CH$_3$OH& (3.332 $\pm$ 0.546)&160.0&2.64--1.71&606.6&4.64--1.36& (5.264 $\pm$ 0.620)&36.42&1.36--2.40&387.2&1.36--4.64\\
 \hline
 Total& &1210.5&&35901.6&& &17144.1&&21785.6&\\\end{tabular}
    \label{tab:Q_Rh_powers}
\end{table*}

\subsection{Change in Radius with Arbitrary Power Law Mass Loss Function}
\label{subsec:2.4}

Here we present solutions with an arbitrary mass loss defined by an inverse power law with an arbitrarily defined power index, $\beta$,
\begin{equation}
    \dot{m} = A \big(r^{-\beta}\big) \,.
    \label{eq:22}
\end{equation}
As discussed in Section~\ref{subsec:2.3}, an inverse power law is a reasonable approximation for the mass loss from gas and dust. The value of $\beta$ is dependent on the molecular species and can be approximated using a power law fit to the production rate as outlined in Section~\ref{sec:03}.

\subsubsection{Circular Orbit}
\label{subsubsec:2.4.1}

For a circular orbit the generic mass loss function assuming an inverse power law is defined as $\dot{m} = A/a^{\beta}$. This mass loss function gives a solution for the final radius of the nucleus of a small body in a circular orbit with a given power law dependency as,
\begin{equation}
     R_f = \bigg[R_0^{3} - \frac{A}{a^{\beta}} \bigg(\frac{h^3}{\mu^2}\frac{3}{4\pi \rho}\bigg) \theta \bigg]^{\frac{1}{3}} \,.
     \label{eq:23}
\end{equation}

\subsubsection{Eccentric Orbit}
\label{subsubsec:2.4.2}

The mass loss function for an eccentric orbit of an arbitrary power law is, 
\begin{equation}
    \dot{m} = A\,\bigg(\,\frac{a(1-e^2)}{1+e\cos{\theta}}\,\bigg)^{-\beta} \,. 
    \label{eq:24}
\end{equation}

The solution of the final radius from perihelion passage through an arbitrary true anomaly for an eccentric orbit is defined as,
\begin{equation}
    \int_{R_0}^{R_f} R^2 \,dR = \bigg(\frac{h^3}{\mu^2}\frac{A}{4\pi \rho}\bigg) \int_{0}^{\theta} \bigg(\frac{a(1-e^2)}{1+e\cos{\theta}}\bigg)^{-\beta} \frac{d \theta}{{(1+e\cos{\theta})^2}} \,.
    \label{eq:25}
\end{equation}

\subsubsection{Hyperbolic Orbit}
\label{subsubsec:2.4.3}

The mass loss for a small body on a hyperbolic trajectory is given as,
\begin{equation}
    \dot{m} = A\,\bigg(\,\frac{a(e^2-1)}{1+e\cos{\theta}}\,\bigg)^{-\beta} \,.
    \label{eq:26}
\end{equation}

Finally, the final nuclear radius over a range of true anomaly for a hyperbolic orbit with an arbitrary power law index is,
\begin{equation}
    \int_{R_0}^{R_f} R^2 \,dR  =  \bigg(\frac{h^3}{\mu^2}\frac{1}{4\pi \rho}\bigg) \int_{0}^{\theta} \bigg(\frac{a(e^2-1)}{1+e\cos{\theta}}\bigg)^{-\beta} \frac{d \theta}{{(1+e\cos{\theta})^2}} \,.
    \label{eq:27}
\end{equation}

\section{Estimates of 3I/ATLAS Production Rates}
\label{sec:03}

In this section we synthesize the production rates of volatiles and refractory compounds detected in the coma of 3I/ATLAS. We collect all published production rates, denoted as $Q_X$, for 3I/ATLAS and provide them in Table~\ref{tab:Q-table} along with the corresponding heliocentric distance, $r_H$, observation date, facility/instrument, and relevant citation. A machine readable version of this table is also provided in the accompanying GitHub \githublink{https://github.com/tfrinck/3IATLAS-PRODUCTON-RATE-TABLE.git}. 

\begin{figure*}
    \centering
    \includegraphics[width=1.\linewidth]{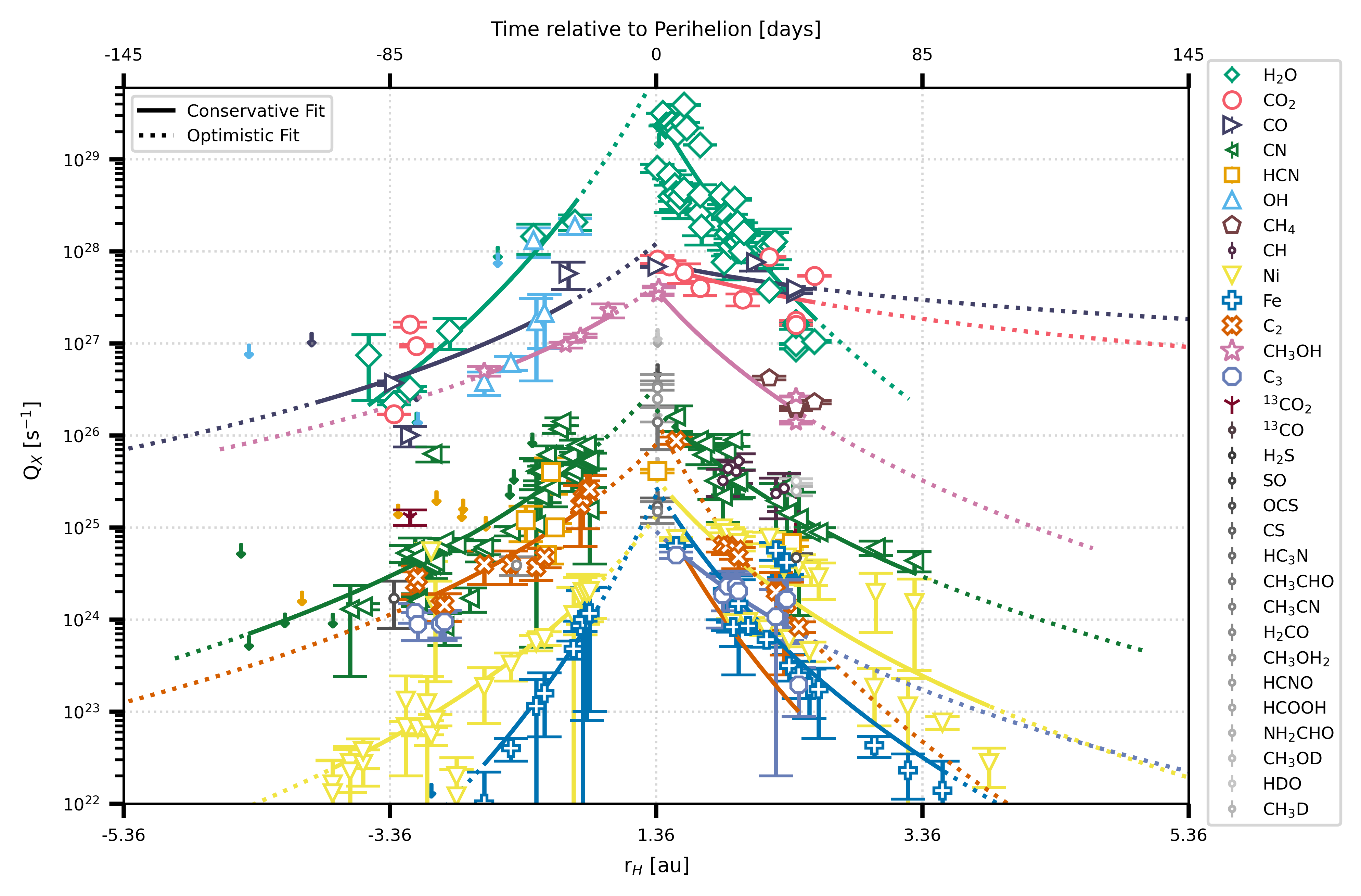}
    \caption{Pre- and post-perihelion production rates for all reported chemical species outgassed by 3I/ATLAS as function of heliocentric distance and time relative to perihelion passage $q$ $\sim$ 1.36 au. Pre-perihelion heliocentric distances are signed to visualize the inbound trajectory. Conservative and optimistic production rate fits are visualized as solid and dotted lines, respectively. Upper limits are indicated as down pointing arrows.}
    \label{fig:Q_date_rhPlot}
\end{figure*}

To ascertain the contribution of each chemical species to the overall activity throughout the trajectory of 3I/ATLAS, we fit the production rates as a function of heliocentric distance, $r_H$, separately for pre- and post-perihelion. We require at least four reported production rates as criteria for fitting for a specific species. In Table~\ref{tab:Q_Rh_powers}, entries with dashes indicate that there was insufficient data to obtain a fit. With that criteria, we are able to fit seven pre-perihelion and nine post-perihelion chemical species. The remaining chemical species detected in the coma of 3I/ATLAS seen in Figure~\ref{fig:Q_date_rhPlot} and Table~\ref{tab:Q-table} had insufficient data to obtain a physically meaningful fit. With this criteria in mind, we maintain that the estimates of mass loss and radius change obtained with these fits are a lower limit. We use the \texttt{scipy.optimize.curve\_fit} which performs a non-linear least-squares to fit the function, 

\begin{equation}
    ln(Q_X) = -\beta ln(r) + b\,.
    \label{eq:28}
\end{equation}
In Equation~(\ref{eq:28}), $\beta$ is the power-law index and $b$ is the intercept \citep{2020SciPy-NMeth}. These parameters are estimated from Equation~(\ref{eq:28}) for the production rate as a function of position. 

For each species we estimate a conservative and optimistic fit for both pre- and post-perihelion measurements. The conservative fits are defined only between the first and last measurement of that species. Optimistic fits are defined over the entire trajectory interior to the relevant sublimation front. In other words, the conservative fits assume 3I/ATLAS was active during the observations, while the optimistic fits assume 3I/ATLAS was active out to large distances. The results of each fit, and resulting mass loss calculated in the next section are provided in Table~\ref{tab:Q_Rh_powers}.

We define the upper limit of the optimistic range of heliocentric distances listed in Table~\ref{tab:Q_Rh_powers} by the sublimation front or nominal distance for transition into gaseous state for each chemical species. The sublimation fronts are derived for the Solar System using a radial temperature profile for protoplanetary disks \citep{Oberg2011}. Sublimation fronts for HCN (and its daughter product CN) and CO, were estimated from sublimation temperatures collected from the NIST Chemistry WebBook \citep{LinstromMallard2001}. For certain chemical species for which sublimation is not the primary mechanism or were not available on the NIST Chemistry WebBook, the upper limit of the optimistic fits were determined based on temperatures cited in literature. We define the optimistic range of r$_H$ for CH$_3$OH by a temperature of evaporation of $\sim$ 130 K \citep{Lippi2024}. For the sublimation of H$_2$O \citet{Kushwaha2025} cited a sublimation temperature of $<$ 170 K. We assume a sublimation temperature of 155 K for H$_2$O and its daughter product OH. C$_2$H$_2$ and C$_2$H$_6$ are potential parent molecules for C$_2$ and are cited to have sublimation temperatures of 90-100 K and 85-95 K, respectively \citep{LinstromMallard2001}. We assume a sublimation temperature of 95 K for C$_2$. Proposed parent molecules for C$_3$ detected in solar system comets include C$_3$H$_4$ \citep{Helbert2005, Stief1972}, C$_3$H$_8$ \citep{Rubin2023}, and other complex hydrocarbons \citep{Biver2024}. \citet{Behmard2019} reported a sublimation temperature of $\sim$ 75 -- 85 K for C$_3$H$_4$; we therefore assume a sublimation temperature of 80 K. The upper limits on optimistic fits for Ni and Fe were derived from estimated sublimation temperatures for the Fe(CO)$_4$ and Ni(CO)$_4$ carbonyls which are 97 -- 108 K and 74 -- 82 K, respectively \citep{Manfroid2021}. We note that the conservative range for H$_2$O we define extends beyond the typical snowline we define of $\sim$ 3.26 au due to the reported H$_2$O production rate at 3.51 au \citep{Xing2025}. We purposefully exclude upper limit values that lie beyond the typical sublimation front for a particular chemical species from the fits and machine readable table. These upper limit production rates remain available for reference in Table~\ref{tab:Q-table}. The upper limits of the range of r$_H$ used for optimistic mass loss estimates (reported in Table~\ref{tab:Q_Rh_powers}) are the positions of the relevant sublimation fronts.

The power-law index parameter, $\beta$, for the conservative and optimistic production rate fits for each chemical species are provided in Table~\ref{tab:Q_Rh_powers} and are arranged by pre- and post-perihelion measurements. We emphasize that the fits are determined with discontinuous observational data and are meant to be interpreted as an approximation of the overall evolution of the production rates throughout the trajectory of 3I/ATLAS. We highlight that the measurements utilized in these fits were obtained with different instruments, observational techniques, and production rate calculation methods. All of which introduced uncertainty in the overall fits as seen in the scatter between contemporaneous measurements in Figure~\ref{fig:Q_date_rhPlot} and uncertainty in computed $\beta$ values.

 Conservative and optimistic fits are plotted in Figure~\ref{fig:Q_date_rhPlot} along with the reported production rates as function of heliocentric distance signed relative to perihelion and days relative to perihelion. The fit to the production rate of H$_2$O has a slightly steeper power law on the inbound trajectory of 3I/ATLAS that is in agreement with other analyses that found a shallower outbound slope \cite{Zhao2026, Tan2026}. Similarly, there is a steeper decreasing slope for post-perihelion C$_2$ confirming estimates made by \citet{Zhao2026}. Overall these calculations produced shallower slopes for CO, CN, Ni, and Fe that are indicative of a more gradual decrease in production as 3I/ATLAS moves out to larger heliocentric distances. Of these asymmetries we note that CO and Fe demonstrate appreciably shallower slopes compared to pre-perihelion production. Asymmetric CO production in pre- and post-perihelion observations of 3I/ATLAS may be indicative of the change from a CO$_2$ \citep{Cordiner2025, Lisse2025a} to CO \citep{Lisse2026,Shinnaka2026, Roth2026} dominated coma. The evolution of this behavior cannot be confirmed without further data, and therefore more conclusive fit to pre-perihelion CO$_2$ production rates. The production of H$_2$O and Fe increased rapidly on the inbound trajectory of 3I/ATLAS whereas CO and CH$_3$OH exhibited a more gradual increase.

\section{ 3I/ATLAS Mass Loss and Radius Change}
\label{sec:04}
\begin{figure}
    \centering
    \includegraphics[width=1.\linewidth]{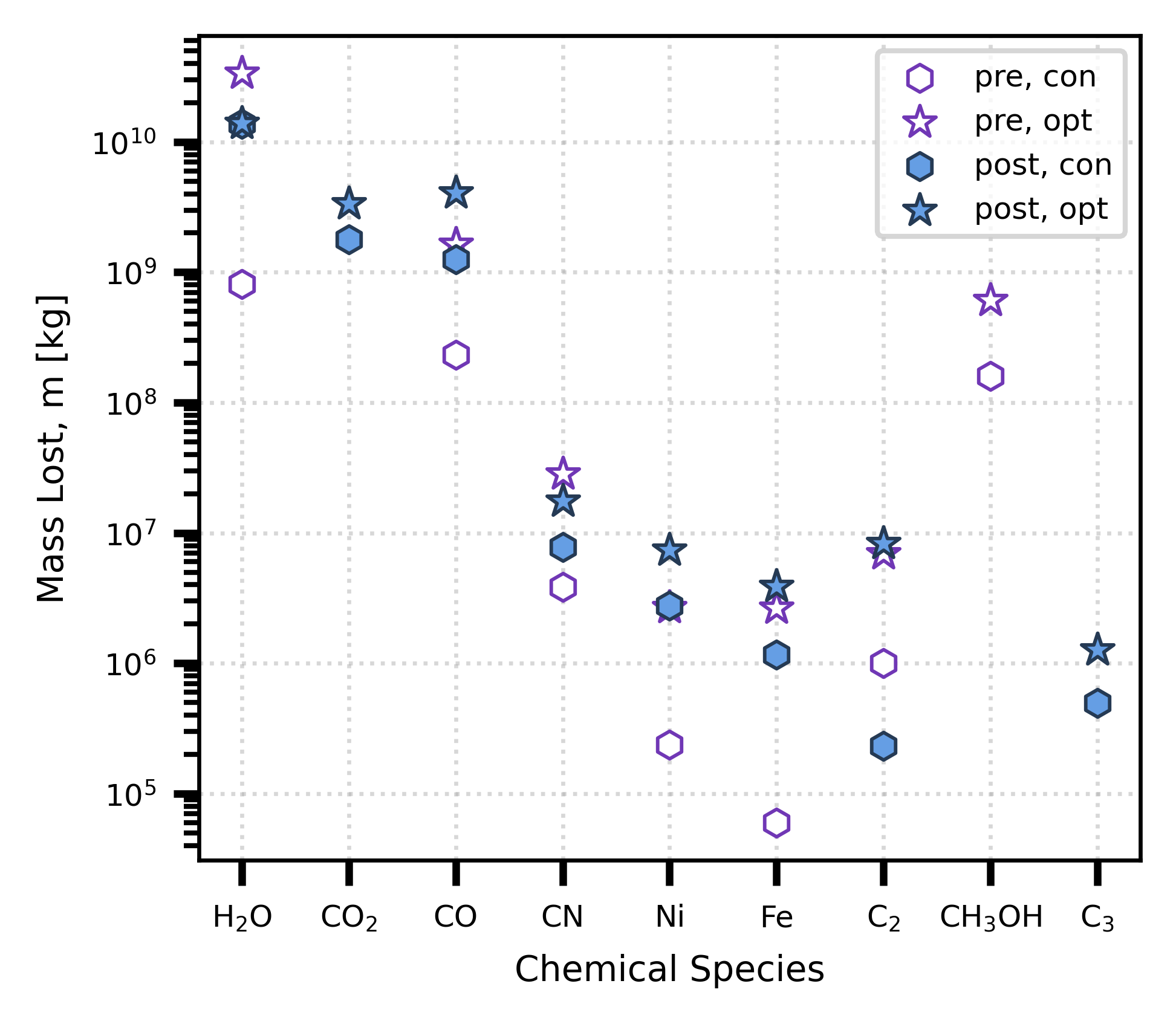}
    \caption{Production of gaseous H$_2$O, CO$_2$, and CO dominated pre- and post-perihelion mass loss. Hexagon markers indicated conservative mass loss estimates. Star markers indicate optimistic mass loss estimates. Unfilled and filled markers are pre- and post-perihelion mass loss estimates, respectively.}
    \label{fig:massloss_species}
\end{figure}

In an effort to contextualize the chemical evolution of the outgassing detected in observations of 3I/ATLAS we estimate the depth down to which the nucleus has eroded. In Subsection~\ref{subsec:4.1} we estimate the mass loss based on the best fitting production rate power laws derived in the previous section. In Subsection~\ref{subsec:4.2} we estimate the corresponding change in radius. 

\subsection{Mass Loss Estimate}
\label{subsec:4.1}

To determine change in nuclear radius we first use the interpolated production rates estimated in Section~\ref{sec:03} to estimate the cumulative mass loss at each position in the trajectory of 3I/ATLAS. We calculate the mass loss rate $\dot{m}$ for each chemical species from pre- and post-perihelion fits to production rates given as,

\begin{equation}
    \dot{m} = Q_X M_X N_A \,,
    \label{eq:29}
\end{equation}
In Equation~(\ref{eq:29}), $Q_X$ is the production rate, $M_X$ is the molar mass, and $N_A$ is Avogadro's number. This provides the mass loss rate in kg s$^{-1}$ at each position over the range of heliocentric distances defined by conservative and optimistic fits for each chemical species. 

We integrate these mass loss rates to calculate conservative and optimistic estimates of the total mass lost from the outgassing from each chemical species and provide those along with the corresponding fit parameters in Table~\ref{tab:Q_Rh_powers}. We estimate the total pre- and post-perihelion mass loss to be $\sim$1.21$\times 10^9$--3.49$\times 10^{10}$ kg and $\sim$(1.71--2.18)$\times 10^{10}$ kg, respectively. The total estimated mass loss from outgassing over the entire trajectory of 3I/ATLAS through the inner Solar System is $\sim$ (1.83--5.77)$\times 10^{10}$ kg. In Figure~\ref{fig:massloss_species} we show all of these mass loss estimates to visualize the contribution of each species to pre- and post-perihelion mass loss. Assuming a density of a typical comet of 0.5 g cm$^{-3}$ and an initial radius of 1.3 $\pm$ 0.2 km estimated for 3I/ATLAS by \citet{Hui2026}, $\sim$ 0.40--1.25$\%$ of the comet's total mass was lost from outgassing. As seen in Figure~\ref{fig:dM_Ro_DRcolorbar}, the mass loss and radius change is strongly dependent on the initial radius of the small body. Estimates of a smaller or larger nuclear radius of 3I/ATLAS have been derived that could increase or decrease the mass and radius loss estimated in this work \citep{Thoss2026, Scarmato2025}. For consistency, we assume 1.3 km as the initial radius of 3I/ATLAS and leave investigation of variable initial radii for future work.

OH and CN are the daughter products of the photodissociation of H$_2$O and HCN in the coma of 3I/ATLAS. We calculate the mass loss excluding OH and HCN production rates to ensure that the mass loss is not overestimated with the inclusion of both parent molecule and daughter product. We include CN rather than HCN due to the spareness of HCN detections throughout the trajectory of 3I/ATLAS. We also acknowledge that the mass loss contribution from radicals such as C$_2$, C$_3$, Fe, and Ni may be an underestimate since we do incorporate their more complex parent molecules that were undetected by extant observations. \citet{McKay2021} measured the abundances of organic compounds detected in the comae of solar system comets and found that C$_2$H$_2$ is a likely parent molecule for C$_2$ but note that it is not the only source. \citet{Manfroid2021} proposed the sublimation or photodissociation of volatile metal bearing carbonyls, Fe(CO)$_4$ and Ni(CO)$_4$, as a possible explanation for the measured Ni/Fe abundances for solar system comets. Ni/Fe abundances have been similarly measured for 3I/ATLAS and 2I/Borisov \citep{Hutsemekers2026a, Guzik2020}. Abundances for these potential parent molecules for solar system comets have been quantified, but we do not attempt to utilize them for 3I/ATLAS without corresponding production rates for the parent molecule. We therefore present the mass loss estimated from these radicals as lower limits. We further acknowledge that we do not account for the molecules typically detected in the comae of solar system comets that were not detected in the coma of 3I/ATLAS. The contribution to the overall mass loss from these undetected species would be relatively insignificant in consideration of H$_2$O, CO, and CO$_2$ dominated activity.

H$_2$O production dominated the mass loss in pre-perihelion estimates while CO, H$_2$O, and CO$_2$ contribute the most to post-perihelion mass loss. Following the production rate asymmetries discussed in Section~\ref{sec:03}, CO shows an evident increase in overall contribution to the total mass loss after perihelion passage.

\subsection{Nucleus Erosion Estimate}
\label{subsec:4.2}

\begin{figure}
    \centering
    \includegraphics[width=1.\linewidth]{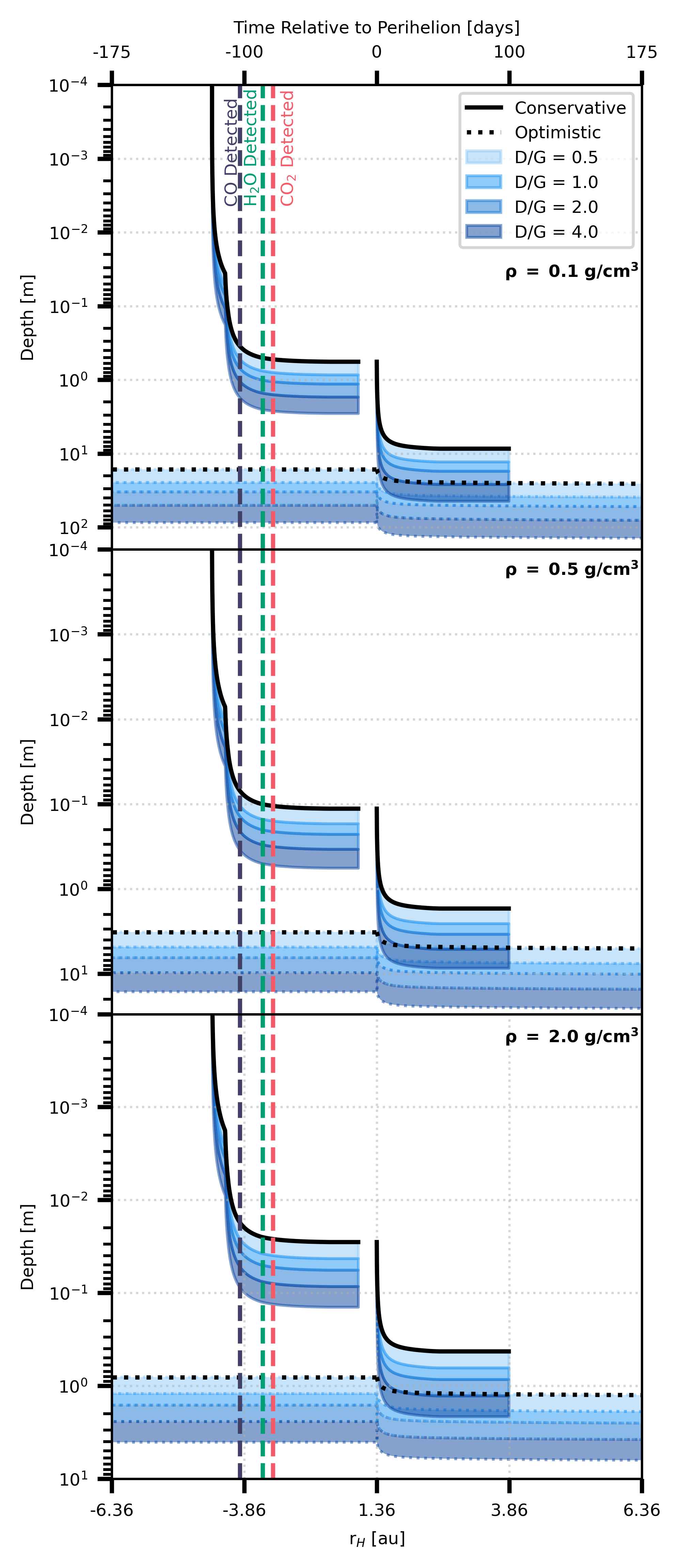}
    \caption{Depth of erosion into 3I/ATLAS as a function of heliocentric distance and time relative to perihelion, where smaller depths are closer to the surface. Each panel assumes a different nucleus density (0.1, 0.5, and 2.0 g/cm$^3$). The conservative and optimistic depth estimates are shown as solid and dotted lines, respectively. The thicker top most lines are from gas alone while shaded regions are estimates considering dust. The vertical dashed lines indicate the heliocentric distances of the first detections of CO, H$_2$O, CO$_2$, respectively.}
    \label{fig:depth_rh}
\end{figure}

We estimate the depth down to which the surface of 3I/ATLAS has been eroded at each position throughout its trajectory. We utilize the mass loss rates calculated in Section~\ref{subsec:4.1} and Equation~(\ref{eq:21}) to determine the final radius for a hyperbolic orbit. For this calculation we assume a typical solar system comet density of 0.5 g cm$^{-3}$ \citep{AHearn2011, Sierks2015, Jorda2016}, initial radius of 1.3 km \citep{Hui2026}, and the mass loss rates over the same conservative and optimistic ranges of heliocentric distance outlined in Table~\ref{tab:Q_Rh_powers}. With these final radii we can determine the change in radius to ascertain the cumulative erosion at each point in the trajectory. This radius change estimate assumes that the contribution of dust to the total mass loss budget was negligible. We later estimate the contribution of dust using various D/G ratios in Subsection~\ref{subsec:4.3}. It is important to note that if the nucleus of 3I/ATLAS had a different bulk density --- for example if it was more or less refractory dominated --- then the resulting change of radius would vary. The density of 0.5 g cm$^{-3}$ encompasses comet nuclei composed of both ices and refractories, whereas the calculations performed in this work primarily consider mass loss from sublimation of ices. We therefore determine the surface erosion of 3I/ATLAS assuming a density, less than low density solar system comets ($<$0.3 g cm$^{-3}$ \citep{Thomas2013}), of 0.1 g cm$^{-3}$ in consideration of a nucleus composed primarily of ices. We also performed the analogous calculation for a nuclear density of 2.0 g cm$^{-3}$ (Figure \ref{fig:depth_rh}).

Conservative and optimistic depth estimates are plotted as solid lines and dotted lines, respectively, in Figure~\ref{fig:depth_rh} as a function of heliocentric distance $r_H$ and time relative to perihelion. Estimates from gas alone are the top most lines at shallower depths while the deeper shaded regions are estimates considering different dust-to-gas (D/G) ratios discussed in Section~\ref{subsec:4.3}. Negative $r_H$ and time indicate the inbound trajectory of 3I/ATLAS and days before perihelion passage, respectively. The depth calculations made with conservative and optimistic mass loss estimates are represented as solid and dotted lines, respectively. Here smaller depths are indicative of less erosion and are closer to the surface or initial radius of 3I/ATLAS. The vertical dashed lines indicate the heliocentric distances at which the first detections of H$_2$O, CO, and CO$_2$ were made from reported observations. To visualize the first detections of these volatiles, the full range of r$_H$ considered for optimistic depth estimates are not shown. 

We find an approximate erosion down to $\sim$0.11--3.24 meters from pre-perihelion outgassing and final depth estimate after post-perihelion outgassing of $\sim$1.71--5.24 meters. These depth estimates comprise 0.13--0.40\% of the assumed 1.3 km initial radius. \citet{Barry-Taylor2026} find a nuclear erosion of 6--70 meters with the inclusion of mass loss from dust, continuous fitting at perihelion, and non-uniform surface activity.

\subsection{Mass Loss and Nucleus Erosion from Dust}
\label{subsec:4.3}

To more accurately account for the mass loss throughout the trajectory of 3I/ATLAS we apply various dust-to-gas ratios (D/G) to our previous mass loss estimates. We apply this as a ratio of the total mass loss rate estimated at each point in the trajectory of 3I/ATLAS. \citet{Moreno2026} presented a comprehensive model of the dust production of 3I/ATLAS from pre- to post-perihelion out to $\sim$ 4 au. The authors find a maximum dust production rate of (0.5--1.8) $\times 10^4$ kg s$^{-1}$ and estimate that the dust-to-gas ratio to be in a range of 1--4. We apply D/G of 0.5, 1.0, 2.0, and 4.0 to the total mass loss rate estimated from each chemical species with a fit (Table~\ref{tab:Q_Rh_powers}) to estimate the mass loss and surface erosion of 3I/ATLAS including dust. We include a D/G of 0.5 outside the cited range for 3I/ATLAS to quantify points in the trajectory at which mass loss may have been dominated by volatiles or dust was ejected by other mechanisms such as thermal cracking, collapsing surface features, etc. \citep{Attree2018, Pajola2017, Choukroun2020, Fulle2019}. We continue to assume an initial radius of 1.3 km. For a D/G = 0.5, we estimate $\sim$ (2.79--8.94)$\times 10^{10}$ kg of mass loss corresponding to $\sim$ 2.57--7.88 meters of erosion. For D/G = 1.0, we estimate $\sim$ $3.72 \times 10^{10}$--$1.19\times 10^{11}$ kg of mass loss and $\sim$ 3.42--10.53 meters of surface erosion. For D/G = 2.0, we calculate $\sim$ $5.58 \times 10^{10}$--$1.79\times 10^{11}$ kg of mass loss and $\sim$ 5.14--15.86 meters of radius change. For D/G = 4.0, we estimate $\sim$ $9.31 \times 10^{10}$--$2.98\times 10^{11}$ kg of mass loss and $\sim$ 8.59--26.65 meters of nuclear erosion. These depth estimates are shown in Figure~\ref{fig:depth_rh} as shaded blue regions. 

\citet{Belyakov2026} estimated a mass loss rate based on a measurement of the dust production rate from observations they obtained with the JWST MIRI on UT 2025 December 15, 16, and 27. With that mass loss rate, \citet{Belyakov2026} similarly calculated 2.5 and 56.5 meters of nuclear erosion for 3I/ATLAS assuming different initial radii and dust particle sizes. We apply these D/G uniformly across the entire range of heliocentric distance defined for the fits but clarify that the D/G is dependent on heliocentric distance. For a more comprehensive picture of the dust loss throughout the trajectory of 3I/ATLAS see \citet{Belyakov2026, Fraser-Gillan2026, Choi2026, Lisse2026, Jewitt2025a, Santana-Ros2025}. 

\section{Discussion}
\label{sec:05}

In this paper, we estimate the mass loss and corresponding change in radius of 3I/ATLAS as a result of the outgassing of chemical species down to a depth of $\sim$ 1.71--5.24 meters. We also estimate the contribution of dust using various D/G and find erosion down to depths $>$10 meters. We provide a method to estimate the change in radius of any small body that is losing mass through a given orbit. We collate a comprehensive table of all available production rates from all chemical species detected in the coma of 3I/ATLAS in Table~\ref{tab:Q-table}. We also provide a full length machine readable version of Table~\ref{tab:Q-table} for the community to reference (available on GitHub \githublink{https://github.com/tfrinck/3IATLAS-PRODUCTON-RATE-TABLE.git}). In addition to erosion estimates, we estimate the mass loss throughout the comet's trajectory and provide a table with the contribution of certain chemical species over different ranges of heliocentric distances. 

Certain fits to production rates found in this work differ from other published analyses largely due to the difference in quantity of data and/or method for obtaining fit parameters. The derived fits do not encapsulate the entire inventory of chemical species in the coma of 3I/ATLAS at each point in the comet's trajectory due to the spareness of the data and chemical species that may have been present but were undetected by extant observations. \citet{BockleeMorvan2017} cite that H$_2$O followed by CO$_2$, CO, CH$_3$OH, CH$_4$ \citep{McKay2021}, H$_2$S, and NH$_3$ \citep{DELLORUSSO2016301} make up the majority of the composition of cometary ices. Most of which were found in the coma of 3I/ATLAS, but with only a few detections (CH$_4$, H$_2$S and NH$_3$). Without further detections, we do not attempt to calculate the mass loss contribution from these and other undetected chemical species. Due to the non-negligible contribution to production rates in solar system comets, we acknowledge that the mass loss, and therefore, radius change estimated in this work for 3I/ATLAS is a lower limit.

We reiterate that the production rates utilized in this work were derived from different instruments with variable observational and reduction techniques. Furthermore, we do not account for the contribution of potential parent molecules of photodissociation products and other undetected chemical species to mass loss or radius change estimates. We therefore present the surface erosion and mass loss calculations as approximations. With these approximations we attempt to contextualize the presence of certain volatile species and the significance of processing down into deeper layers of the nucleus. 

The activity for most solar system comets is primarily driven by the sublimation of H$_2$O, CO, and CO$_2$ ices for which abundances provide insights into the formation of the early Solar System \citep{Jewitt2009}. Of particular interest are CO and CO$_2$ which drive activity of comets at relatively large heliocentric distances due to their low sublimation temperatures. \citet{Harrington-Pinto2022} presented a survey compiling production rates of CO and CO$_2$ for 25 solar system comets and categorized them based on which volatile dominates the activity. \citet{Harrington-Pinto2022} demonstrated that dynamically new LPCs, upon their first perihelion passage, produce more CO$_2$ than CO compared to dynamically old LPCs after several visits to the inner Solar System, consistent with previous smaller surveys of solar system comets \citep{AHearn2012, Ootsubo2012}. \citet{Harrington-Pinto2022} posited that CO is depleted in dynamically new LPCs from galactic cosmic rays altering the surface layers of ices resulting in an overall higher CO$_2$/CO mixing ratios during their first passage in the inner Solar System \citep{Gronoff2020, Maggiolo2020}. This is consistent with observations of 3I/ATLAS that measured a CO$_2$ enriched pre-perihelion coma \citep{Cordiner2025, Lisse2025a}.

The energy from galactic cosmic rays (GCR) can be deposited down to a depth of $\sim$70 meters into the surface of a comet depending on the bulk properties of the comet and the energy of the GCRs \citet{Gronoff2020}. \citet{Maggiolo2020} argued that sufficiently high energy GCRs alter the chemical composition of cometary surface ices via radiolysis of H$_2$O, CO, and CO$_2$ down to $\sim$10--20 meters. In a similar analysis for 3I/ATLAS, \citet{Maggiolo2025} estimated GCR conversion of CO to CO$_2$ down to a depth of $\sim$15--20 meters to explain the unusually large CO$_2$/CO mixing ratio measured with \textit{JWST} pre-perihelion observations \citep{Cordiner2025}. \citet{Maggiolo2025} also estimated pre-perihelion erosion of the surface of 3I/ATLAS to be less than $\sim$ 1 meter and up to a few tens of meters near perihelion passage. Therefore, \citet{Maggiolo2025} concluded that it is unlikely the erosion of the surface of 3I/ATLAS would reveal deeper layers of pristine material. The erosion of the nucleus we estimate of $\sim$1.71--5.24 meters from gas alone is consistent with these previous estimates. With the inclusion of mass lost from dust down to depths $>$10 meters, later post-perihelion observations of 3I/ATLAS may be representative of more pristine layers. In the future, observations of deeper layers of an interstellar comet may reveal more primordial compositions. This may be possible for an interstellar object with a trajectory that spends more time in the inner Solar System or smaller initial radius.

\section{Conclusions}
\label{sec:06}

Here we summarize the main results presented in this paper:

\begin{enumerate}
    \item Equations~(\ref{eq:10}), (\ref{eq:12}), (\ref{eq:14}), (\ref{eq:23}), (\ref{eq:25}), and (\ref{eq:27}) provide a methodology to estimate the final nuclear radius of a small body undergoing mass loss for an arbitrary trajectory and mass loss function. 
    \item We compiled all measured production rates of 3I/ATLAS and provide a table of these measurements and the associated heliocentric distances, observation dates, facility/instrument, and citation.
    \item We perform conservative and optimistic fits of observed production rates as a function of pre- and post-perihelion positions of 3I/ATLAS. We fit production rates for seven and nine different chemical species detected in the coma from pre- and post-perihelion observations, respectively.
    \item The total estimated mass loss from gas alone for 3I/ATLAS based on fits to production rates is $\sim$10$^{10}$ kg.
    \item Assuming an initial radius of $\sim$1.3 km, we approximate the erosion of the surface of 3I/ATLAS to be down to a depth of $\sim$1.71--5.24 meters.
    \item In consideration of dust-to-gas ratios in range 1--4 we estimate erosion of the surface to be $>$10 meters.
    \item Based on mass loss and radius change estimates determined from the contribution of dust and gas, we conclude that it is possible that the surface has been eroded down to layers of unaltered material. In this work, pre-perihelion/post-perihelion depths are estimated to be $\sim$0.11-3.24/1.71-5.24  meters from gas and down to $\sim$0.57-16.36/8.59--26.65 meters from the most optimistic D/G of 4.0. 
\end{enumerate}

\section*{Acknowledgements}

We thank the anonymous reviewer for their insightful comments and suggestions that strengthened the scientific content of this manuscript.

DZS acknowledges funding support from JWST GO 5959, which was provided by NASA through a grant from the Space Telescope Science Institute. We thank Adina Feinstein, Cassidy Walker, Aster Barry-Taylor, and Martin Cordiner for useful conversations. 

\textit{Software}: \texttt{matplotlib} \citep{Hunter:2007}, \texttt{numpy} \citep{harris2020array}, \texttt{scipy} \citep{2020SciPy-NMeth}.

\section*{Data Availability}

The machine readable table of production rates utilized in this publication is available at this link: \href{https://github.com/tfrinck/3IATLAS-PRODUCTON-RATE-TABLE.git}{\textcolor{black}\faGithub}.

\bibliographystyle{mnras}
\bibliography{example} 

@ARTICLE{Fitzsimmons2018,
       author = {{Fitzsimmons}, Alan and {Snodgrass}, Colin and {Rozitis}, Ben and {Yang}, Bin and {Hyland}, M{\'e}abh and {Seccull}, Tom and {Bannister}, Michele T. and {Fraser}, Wesley C. and {Jedicke}, Robert and {Lacerda}, Pedro},
        title = "{Spectroscopy and thermal modelling of the first interstellar object 1I/2017 U1 `Oumuamua}",
      journal = {Nature Astronomy},
         year = 2018,
        month = dec,
       volume = {2},
        pages = {133-137},
          doi = {10.1038/s41550-017-0361-4},
archivePrefix = {arXiv},
       eprint = {1712.06552},
 primaryClass = {astro-ph.EP},
       adsurl = {https://ui.adsabs.harvard.edu/abs/2018NatAs...2..133F}
}

@ARTICLE{Taylor2025a,
       author = {{Taylor}, Aster G. and {Seligman}, Darryl Z.},
        title = "{The Kinematic Age of 3I/ATLAS and Its Implications for Early Planet Formation}",
      journal = {\apjl},
         year = 2025,
        month = sep,
       volume = {990},
       number = {1},
          eid = {L14},
        pages = {L14},
          doi = {10.3847/2041-8213/adfa28},
archivePrefix = {arXiv},
       eprint = {2507.08111},
 primaryClass = {astro-ph.EP},
       adsurl = {https://ui.adsabs.harvard.edu/abs/2025ApJ...990L..14T}
}

@ARTICLE{Lisse2025a,
       author = {{Lisse}, Carey M. and {Bach}, Yoonsoo P. and {Bryan}, Sean A. and {Crill}, Brendan P. and {Korngut}, Phil M. and {Cukierman}, Ari J. and {Werner}, Michael W. and {Cooray}, Asantha and {Zemcov}, Michael and {Tolls}, Volker and {Melnick}, Gary J. and {Faisst}, Andreas L. and {Dowell}, C. Darren and {Choi}, Seungwon and {Geem}, Jooyeon and {Ishiguro}, Masateru and {Jo}, Hangbin and {Lim}, Bumhoo and {Mahlke}, Max and {Hora}, Joseph L. and {Cheng}, Yun-Ting and {Everett}, Spencer and {Lee}, Jeong-Eun and {Rustamkulov}, Zafar and {Jin}, Sunho and {Hui}, Howard and {Masters}, Daniel C. and {Nguyen}, Chi H. and {Paladini}, Roberta and {Yang}, Yujin and {Bock}, James J. and {Dor{\'e}}, O. and {Sitko}, M.~L. and {Champagne}, C. and {Connelley}, M. and {Emery}, J.~P. and {Fernandez}, Y.~R. and {Reach}, W.~T.},
        title = "{SPHEREx Pre-Perihelion Mapping of $\mathrm{H_2O}$, $\mathrm{CO_2}$, and $\mathrm{CO}$ in Interstellar Object 3I/ATLAS}",
      journal = {arXiv e-prints},
         year = 2025,
        month = dec,
          eid = {arXiv:2512.07318},
        pages = {arXiv:2512.07318},
archivePrefix = {arXiv},
       eprint = {2512.07318},
 primaryClass = {astro-ph.EP},
       adsurl = {https://ui.adsabs.harvard.edu/abs/2025arXiv251207318L}
}

@ARTICLE{Cordiner2025,
       author = {{Cordiner}, Martin A. and {Roth}, Nathan X. and {Kelley}, Michael S.~P. and {Bodewits}, Dennis and {Charnley}, Steven B. and {Drozdovskaya}, Maria N. and {Farnocchia}, Davide and {Micheli}, Marco and {Milam}, Stefanie N. and {Opitom}, Cyrielle and {Schwamb}, Megan E. and {Thomas}, Cristina A. and {Bagnulo}, Stefano},
        title = "{JWST Detection of a Carbon-dioxide-dominated Gas Coma Surrounding Interstellar Object 3I/ATLAS}",
      journal = {\apjl},
         year = 2025,
        month = oct,
       volume = {991},
       number = {2},
          eid = {L43},
        pages = {L43},
          doi = {10.3847/2041-8213/ae0647},
archivePrefix = {arXiv},
       eprint = {2508.18209},
 primaryClass = {astro-ph.EP},
       adsurl = {https://ui.adsabs.harvard.edu/abs/2025ApJ...991L..43C}
}

@ARTICLE{Maggiolo2025,
       author = {{Maggiolo}, R. and {Dhooghe}, F. and {Gronoff}, G. and {de Keyser}, J. and {Cessateur}, G.},
        title = "{Interstellar Comet 3I/ATLAS: Evidence for Galactic Cosmic Ray Processing}",
      journal = {arXiv e-prints},
         year = 2025,
        month = oct,
          eid = {arXiv:2510.26308},
        pages = {arXiv:2510.26308},
          doi = {10.48550/arXiv.2510.26308},
archivePrefix = {arXiv},
       eprint = {2510.26308},
 primaryClass = {astro-ph.EP},
       adsurl = {https://ui.adsabs.harvard.edu/abs/2025arXiv251026308M}
}

@ARTICLE{Tonry2018a,
       author = {{Tonry}, J.~L. and {Denneau}, L. and {Heinze}, A.~N. and {Stalder}, B. and {Smith}, K.~W. and {Smartt}, S.~J. and {Stubbs}, C.~W. and {Weiland}, H.~J. and {Rest}, A.},
        title = "{ATLAS: A High-cadence All-sky Survey System}",
      journal = {\pasp},
         year = 2018,
        month = jun,
       volume = {130},
       number = {988},
        pages = {064505},
          doi = {10.1088/1538-3873/aabadf},
archivePrefix = {arXiv},
       eprint = {1802.00879},
 primaryClass = {astro-ph.IM},
       adsurl = {https://ui.adsabs.harvard.edu/abs/2018PASP..130f4505T}
}

@ARTICLE{Denneau2025,
       author = {{Denneau}, L. and {Siverd}, R. and {Tonry}, J. and {Weiland}, H. and {Erasmus}, N. and {Fitzsimmons}, A. and {Robinson}, J.},
        title = "{3I/ATLAS = C/2025 N1 (ATLAS)}",
      journal = {MPEC},
         year = 2025,
        month = jul,
       number = {2025-N12}
}

@ARTICLE{Seligman2025a,
       author = {{Seligman}, Darryl Z. and {Micheli}, Marco and {Farnocchia}, Davide and {Denneau}, Larry and {Noonan}, John W. and {Hsieh}, Henry H. and {Santana-Ros}, Toni and {Tonry}, John and {Auchettl}, Katie and {Conversi}, Luca and {Devog{\`e}le}, Maxime and {Faggioli}, Laura and {Feinstein}, Adina D. and {Fenucci}, Marco and {Ferrais}, Marin and {Frincke}, Tessa and {Gillon}, Michael and {Hainaut}, Olivier R. and {Hart}, Kyle and {Hoffman}, Andrew and {Holt}, Carrie E. and {Hoogendam}, Willem B. and {Huber}, Mark E. and {Jehin}, Emmanuel and {Kareta}, Theodore and {Keane}, Jacqueline V. and {Kelley}, Michael S.~P. and {Lister}, Tim and {Mandt}, Kathleen and {Manfroid}, Jean and {Mar{\v{c}}eta}, Du{\v{s}}an and {Meech}, Karen J. and {Amine Miftah}, Mohamed and {Morgan}, Marvin and {Oca{\~n}a}, Francisco and {Pe{\~n}a-Asensio}, Eloy and {Shappee}, Benjamin J. and {Siverd}, Robert J. and {Taylor}, Aster G. and {Tucker}, Michael A. and {Wainscoat}, Richard and {Weryk}, Robert and {Wray}, James J. and {Yaginuma}, Atsuhiro and {Yang}, Bin and {Ye}, Quanzhi and {Zhang}, Qicheng},
        title = "{Discovery and Preliminary Characterization of a Third Interstellar Object: 3I/ATLAS}",
      journal = {\apjl},
         year = 2025,
        month = aug,
       volume = {989},
       number = {2},
          eid = {L36},
        pages = {L36},
          doi = {10.3847/2041-8213/adf49a},
archivePrefix = {arXiv},
       eprint = {2507.02757},
 primaryClass = {astro-ph.EP},
       adsurl = {https://ui.adsabs.harvard.edu/abs/2025ApJ...989L..36S}
}

@ARTICLE{Choi2026,
       author = {{Choi}, Seungwon and {Ishiguro}, Masateru and {Takahashi}, Jun and {Saito}, Tomoki and {Bach}, Yoonsoo P. and {Lim}, Bumhoo and {Naito}, Hiroyuki and {Geem}, Jooyeon and {Jin}, Sunho and {Seo}, Jinguk and {Ju}, Hyeonwoo and {Akitakya}, Hiroshi and {Kawabata}, Koji S. and {Sasada}, Mahito and {Doi}, Kazuya and {Kubota}, Hisayuki and {Takagi}, Seiko and {Watanabe}, Makoto and {Sekiguchi}, Tomohiko and {Im}, Myungshin},
        title = "{Dust Properties of the Interstellar Object 3I/ATLAS Revealed by Optical and Near-Infrared Polarimetry}",
      journal = {arXiv e-prints},
         year = 2026,
        month = jan,
          eid = {arXiv:2601.08591},
        pages = {arXiv:2601.08591},
          doi = {10.48550/arXiv.2601.08591},
archivePrefix = {arXiv},
       eprint = {2601.08591},
 primaryClass = {astro-ph.EP},
       adsurl = {https://ui.adsabs.harvard.edu/abs/2026arXiv260108591C}
}

@ARTICLE{Coulson2025,
       author = {{Coulson}, Iain M. and {Kuan}, Yi-Jehng and {Charnley}, Steven B. and {Cordiner}, Martin A. and {Chuang}, Yo-Ling and {Lee}, Yueh-Ning and {Lin}, Min-Kai and {Milam}, Stefanie N. and {Pimpanuwat}, Bannawit and {Roth}, Nathan X. and {{\.Z}{\'o}{\l}towski}, Micha{\l}},
        title = "{JCMT detection of HCN emission from 3I/ATLAS at 2.1 AU}",
      journal = {arXiv e-prints},
         year = 2025,
        month = oct,
          eid = {arXiv:2510.02817},
        pages = {arXiv:2510.02817},
          doi = {10.48550/arXiv.2510.02817},
archivePrefix = {arXiv},
       eprint = {2510.02817},
 primaryClass = {astro-ph.EP},
       adsurl = {https://ui.adsabs.harvard.edu/abs/2025arXiv251002817C}
}

@ARTICLE{Combi2025,
       author = {{Combi}, M.~R. and {M{\^a}kinen}, T. and {Bertaux}, J.-L. and {Quemerais}, E. and {Ferron}, S. and {Lallement}, R. and {Schmidt}, W.},
        title = "{Water Production of Interstellar Comet 3I/ATLAS from SOHO/SWAN Observations after Perihelion}",
      journal = {arXiv e-prints},
         year = 2025,
        month = dec,
          eid = {arXiv:2512.22354},
        pages = {arXiv:2512.22354},
          doi = {10.48550/arXiv.2512.22354},
archivePrefix = {arXiv},
       eprint = {2512.22354},
 primaryClass = {astro-ph.EP},
       adsurl = {https://ui.adsabs.harvard.edu/abs/2025arXiv251222354C}
}

@ARTICLE{Tan2026,
       author = {{Tan}, Hanjie and {Yan}, Xiaoran and {Li}, Jian-Yang},
        title = "{Perihelion Asymmetry in the Water Production Rate of the Interstellar Object 3I/ATLAS}",
      journal = {arXiv e-prints},
         year = 2026,
        month = jan,
          eid = {arXiv:2601.15443},
        pages = {arXiv:2601.15443},
archivePrefix = {arXiv},
       eprint = {2601.15443},
 primaryClass = {astro-ph.EP},
       adsurl = {https://ui.adsabs.harvard.edu/abs/2026arXiv260115443T}
}

@ARTICLE{Hoogendam2026,
       author = {{Hoogendam}, Willem B. and {Jones}, David O. and {Yang}, Bin and {Shappee}, Benjamin J. and {Wray}, James J. and {Meech}, Karen J. and {Ashall}, Christopher and {Desai}, Dhvanil D. and {Hinkle}, Jason T. and {Hoffman}, Andrew M. and {Medler}, Kyle and {Pfeffer}, Cameron and {Zhao}, Ruining},
        title = "{Post-Perihelion Integral Field Spectroscopy of the Interstellar Comet 3I/ATLAS}",
      journal = {arXiv e-prints},
         year = 2026,
        month = jan,
          eid = {arXiv:2601.16983},
        pages = {arXiv:2601.16983},
archivePrefix = {arXiv},
       eprint = {2601.16983},
 primaryClass = {astro-ph.EP},
       adsurl = {https://ui.adsabs.harvard.edu/abs/2026arXiv260116983H}
}

@ARTICLE{Lisse2026,
       author = {{Lisse}, C.~M. and {Bach}, YP. and {Bryan}, S.~A. and {Korngut}, P.~M. and {Crill}, B.~P. and {Cukierman}, A.~J. and {Dor{\'e}}, O. and {Cooray}, A. and {Fabinsky}, B. and {Faisst}, A.~L. and {Hui}, H. and {Melnick}, G.~J. and {Nguyen}, C.~H. and {Rustamkulov}, Z. and {Tolls}, V. and {Werner}, M.~W.},
        title = "{SPHEREx Re-Observation of Interstellar Object 3I/ATLAS in December 2025: Detection of Increased Post-Perihelion Activity, Refractory Coma Dust, and New Coma Gas Species}",
      journal = {arXiv e-prints},
         year = 2026,
        month = jan,
          eid = {arXiv:2601.06759},
        pages = {arXiv:2601.06759},
          doi = {10.48550/arXiv.2601.06759},
archivePrefix = {arXiv},
       eprint = {2601.06759},
 primaryClass = {astro-ph.EP},
       adsurl = {https://ui.adsabs.harvard.edu/abs/2026arXiv260106759L}
}

@ARTICLE{Li2026,
       author = {{Li}, Juncen and {Shi}, Xian and {Hui}, Man-To and {Shi}, Jianchun},
        title = "{Pre-perihelion Volatile Evolution of Interstellar Comet 3I/ATLAS Indicating Significant Contribution from Extended Source in the Coma}",
      journal = {arXiv e-prints},
         year = 2026,
        month = feb,
          eid = {arXiv:2602.14218},
        pages = {arXiv:2602.14218},
          doi = {10.48550/arXiv.2602.14218},
archivePrefix = {arXiv},
       eprint = {2602.14218},
 primaryClass = {astro-ph.EP},
       adsurl = {https://ui.adsabs.harvard.edu/abs/2026arXiv260214218L}
}

@ARTICLE{Paek2026,
       author = {{Paek}, Gregory S.~H. and {Im}, Myungshin and {Jeong}, Mankeun and {Choi}, Hyeonho and {Bach}, Yoonsoo P. and {Ishiguro}, Masateru and {Lim}, Bumhoo and {Chang}, Seo-Won and {Kim}, Ji Hoon and {Geem}, Jooyeon and {Hoogendam}, Willem B.},
        title = "{Pre-perihelion Emergence of the CN Gas Coma in 3I/ATLAS Temporally and Spatially Resolved by the 7-Dimensional Telescope}",
      journal = {arXiv e-prints},
         year = 2026,
        month = feb,
          eid = {arXiv:2602.12930},
        pages = {arXiv:2602.12930},
          doi = {10.48550/arXiv.2602.12930},
archivePrefix = {arXiv},
       eprint = {2602.12930},
 primaryClass = {astro-ph.EP},
       adsurl = {https://ui.adsabs.harvard.edu/abs/2026arXiv260212930P}
}

@ARTICLE{Belyakov2026,
       author = {{Belyakov}, Matthew and {Wong}, Ian and {Bolin}, Bryce T. and {Ryleigh Davis}, M. and {Bromley}, Steven J. and {Lisse}, Carey M. and {Brown}, Michael E.},
        title = "{The Volatile Inventory of 3I/ATLAS as seen with JWST/MIRI}",
      journal = {arXiv e-prints},
         year = 2026,
        month = jan,
          eid = {arXiv:2601.22034},
        pages = {arXiv:2601.22034},
          doi = {10.48550/arXiv.2601.22034},
archivePrefix = {arXiv},
       eprint = {2601.22034},
 primaryClass = {astro-ph.EP},
       adsurl = {https://ui.adsabs.harvard.edu/abs/2026arXiv260122034B}
}

@ARTICLE{Jewitt2025a,
       author = {{Jewitt}, David and {Hui}, Man-To and {Mutchler}, Max and {Kim}, Yoonyoung and {Agarwal}, Jessica},
        title = "{Hubble Space Telescope Observations of the Interstellar Interloper 3I/ATLAS}",
      journal = {\apjl},
         year = 2025,
        month = sep,
       volume = {990},
       number = {1},
          eid = {L2},
        pages = {L2},
          doi = {10.3847/2041-8213/adf8d8},
archivePrefix = {arXiv},
       eprint = {2508.02934},
 primaryClass = {astro-ph.EP},
       adsurl = {https://ui.adsabs.harvard.edu/abs/2025ApJ...990L...2J}
}

@ARTICLE{Hui2026,
       author = {{Hui}, Man-To and {Jewitt}, David and {Mutchler}, Max J. and {Agarwal}, Jessica and {Kim}, Yoonyoung},
        title = "{Nucleus and Postperihelion Activity of Interstellar Object 3I/ATLAS Observed by Hubble Space Telescope}",
      journal = {arXiv e-prints},
         year = 2026,
        month = jan,
          eid = {arXiv:2601.21569},
        pages = {arXiv:2601.21569},
          doi = {10.48550/arXiv.2601.21569},
archivePrefix = {arXiv},
       eprint = {2601.21569},
 primaryClass = {astro-ph.EP},
       adsurl = {https://ui.adsabs.harvard.edu/abs/2026arXiv260121569H}
}

@ARTICLE{Hoogendam2025b,
       author = {{Hoogendam}, W.~B. and {Kuesters}, D. and {Shappee}, B.~J. and {Aldering}, G. and {Wray}, J.~J. and {Yang}, B. and {Meech}, K.~J. and {Tucker}, M.~A. and {Huber}, M.~E. and {Auchettl}, K. and {Angus}, C.~R. and {Desai}, D.~D. and {Hinkle}, J.~T. and {Kiyokawa}, J. and {Paek}, G.~S.~H. and {Romagnoli}, S. and {Shi}, J. and {Syncatto}, A. and {Ashall}, C. and {Dixon}, M. and {Hart}, K. and {Hoffman}, A.~M. and {Jones}, D.~O. and {Medler}, K. and {Pfeffer}, C.},
        title = "{University of Hawaii 88-inch Telescope Observations of the Interstellar Comet 3I/ATLAS: Spectrophotometric Blue-Sensitive Spectral Time Series Spanning Two Months from Discovery}",
      journal = {arXiv e-prints},
         year = 2025,
        month = dec,
          eid = {arXiv:2512.09020},
        pages = {arXiv:2512.09020},
          doi = {10.48550/arXiv.2512.09020},
archivePrefix = {arXiv},
       eprint = {2512.09020},
 primaryClass = {astro-ph.EP},
       adsurl = {https://ui.adsabs.harvard.edu/abs/2025arXiv251209020H}
}

@ARTICLE{Hinkle2025,
       author = {{Hinkle}, Jason T. and {Yang}, Bin and {Meech}, Karen J. and {Hoffman}, Andrew and {Shappee}, Benjamin J. and {Hoogendam}, W.~B. and {Wray}, James J.},
        title = "{JCMT Constraints on the Early-Time HCN and CO Emission and HCN Temporal Evolution of 3I/ATLAS}",
      journal = {arXiv e-prints},
         year = 2025,
        month = dec,
          eid = {arXiv:2512.02106},
        pages = {arXiv:2512.02106},
          doi = {10.48550/arXiv.2512.02106},
archivePrefix = {arXiv},
       eprint = {2512.02106},
 primaryClass = {astro-ph.EP},
       adsurl = {https://ui.adsabs.harvard.edu/abs/2025arXiv251202106H}
}

@ARTICLE{Roth2025,
       author = {{Roth}, Nathan X. and {Cordiner}, Martin A. and {Bockel{\'e}e-Morvan}, Dominique and {Biver}, Nicolas and {Crovisier}, Jacques and {Milam}, Stefanie N. and {Lellouch}, Emmanuel and {Santos-Sanz}, Pablo and {Lis}, Dariusz C. and {Qi}, Chunhua and {Foster}, K.~D. and {Boissier}, J{\'e}r{\'e}mie and {Furuya}, Kenji and {Moreno}, Raphael and {Charnley}, Steven B. and {Remijan}, Anthony J. and {Kuan}, Yi-Jehng and {Hart}, Lillian X.},
        title = "{CH$_3$OH and HCN in Interstellar Comet 3I/ATLAS Mapped with the ALMA Atacama Compact Array: Distinct Outgassing Behaviors and a Remarkably High CH$_3$OH/HCN Production Rate Ratio}",
      journal = {arXiv e-prints},
         year = 2025,
        month = nov,
          eid = {arXiv:2511.20845},
        pages = {arXiv:2511.20845},
          doi = {10.48550/arXiv.2511.20845},
archivePrefix = {arXiv},
       eprint = {2511.20845},
 primaryClass = {astro-ph.EP},
       adsurl = {https://ui.adsabs.harvard.edu/abs/2025arXiv251120845R}
}

@ARTICLE{Medler2026,
       author = {{Medler}, Kyle and {Hoogendam}, Willem B. and {Ashall}, Christopher and {Yang}, Bin and {Wray}, James J. and {Shappee}, Benjamin J. and {Meech}, Karen J. and {Tucker}, Michael A. and {Auchettl}, Katie and {Desai}, Dhvanil D. and {Hinkle}, Jason T. and {Hoffman}, Andrew M. and {Huber}, Mark E. and {Jones}, David O. and {Zhao}, Ruining},
        title = "{Contemporaneous Optical and Near-Infrared Observations of the Interstellar Comet 3I/ATLAS Pre- and Post-Perihelion}",
      journal = {arXiv e-prints},
         year = 2026,
        month = feb,
          eid = {arXiv:2602.23586},
        pages = {arXiv:2602.23586},
          doi = {10.48550/arXiv.2602.23586},
archivePrefix = {arXiv},
       eprint = {2602.23586},
 primaryClass = {astro-ph.EP},
       adsurl = {https://ui.adsabs.harvard.edu/abs/2026arXiv260223586M}
}

@ARTICLE{Hoogendam2025a,
       author = {{Hoogendam}, W.~B. and {Shappee}, B.~J. and {Wray}, J.~J. and {Yang}, B. and {Meech}, K.~J. and {Ashall}, C. and {Desai}, D.~D. and {Hart}, K. and {Hinkle}, J.~T. and {Hoffman}, A. and {Hu}, E.~M. and {Jones}, D.~O. and {Medler}, K. and {Pfeffer}, C.},
        title = "{Spatial Profiles of 3I/ATLAS CN and Ni Outgassing from Keck/KCWI Integral Field Spectroscopy}",
      journal = {arXiv e-prints},
         year = 2025,
        month = oct,
          eid = {arXiv:2510.11779},
        pages = {arXiv:2510.11779},
          doi = {10.48550/arXiv.2510.11779},
archivePrefix = {arXiv},
       eprint = {2510.11779},
 primaryClass = {astro-ph.EP},
       adsurl = {https://ui.adsabs.harvard.edu/abs/2025arXiv251011779H}
}

@ARTICLE{Hutsemekers2026a,
       author = {{Hutsem{\'e}kers}, Damien and {Manfroid}, Jean and {Jehin}, Emmanu{\"e}l and {Opitom}, Cyrielle and {Rahatgaonkar}, Rohan and {Bannister}, Michele and {Carvajal}, Juan Pablo and {Dorsey}, Rosemary and {Krishnakumar}, Aravind and {Luco}, Baltasar and {Murphy}, Brian and {Puzia}, Thomas H.},
        title = "{Pre-perihelion evolution of the NiI/FeI abundance ratio in the coma of the interstellar comet 3I/ATLAS: From extreme to normal}",
      journal = {\aap},
         year = 2026,
        month = jan,
       volume = {706},
          eid = {A43},
        pages = {A43},
          doi = {10.1051/0004-6361/202557484},
archivePrefix = {arXiv},
       eprint = {2509.26053},
 primaryClass = {astro-ph.EP},
       adsurl = {https://ui.adsabs.harvard.edu/abs/2026A&A...706A..43H}
}

@ARTICLE{Ye2025,
       author = {{Ye}, Quanzhi and {Kelley}, Michael S.~P. and {Hsieh}, Henry H. and {Bellm}, Eric C. and {Chen}, Tracy X. and {Dekany}, Richard and {Drake}, Andrew and {Groom}, Steven L. and {Helou}, George and {Kulkarni}, Shrinivas R. and {Prince}, Thomas A. and {Riddle}, Reed},
        title = "{Prediscovery Activity of New Interstellar Object 3I/ATLAS: Rapid Brightening from 6 to 4 au}",
      journal = {\apjl},
         year = 2025,
        month = nov,
       volume = {993},
       number = {1},
          eid = {L31},
        pages = {L31},
          doi = {10.3847/2041-8213/ae147b},
archivePrefix = {arXiv},
       eprint = {2509.08792},
 primaryClass = {astro-ph.EP},
       adsurl = {https://ui.adsabs.harvard.edu/abs/2025ApJ...993L..31Y}
}

@ARTICLE{Tonry2025,
       author = {{Tonry}, John L. and {Denneau}, Jr., Larry and {Alarc{\'o}n}, Miguel R. and {Clocchiatti}, Alejandro and {Erasmus}, Nicolas and {Fitzsimmons}, Alan and {Licandro}, Javier and {Meech}, Karen J. and {Siverd}, Robert and {Weiland}, Henry},
        title = "{ATLAS Photometry of Interstellar Object 3I/ATLAS}",
      journal = {\apjl},
         year = 2025,
        month = dec,
       volume = {995},
       number = {1},
          eid = {L15},
        pages = {L15},
          doi = {10.3847/2041-8213/ae1f12},
archivePrefix = {arXiv},
       eprint = {2509.05562},
 primaryClass = {astro-ph.EP},
       adsurl = {https://ui.adsabs.harvard.edu/abs/2025ApJ...995L..15T}
}

@ARTICLE{Salazar-Manzano2025,
       author = {{Salazar Manzano}, Luis E. and {Lin}, Hsing Wen and {Taylor}, Aster G. and {Seligman}, Darryl Z. and {Adams}, Fred C. and {Gerdes}, David W. and {Ruch}, Thomas and {Frincke}, Tessa T. and {Napier}, Kevin J.},
        title = "{Onset of CN Emission in 3I/ATLAS: Evidence for Strong Carbon-chain Depletion}",
      journal = {\apjl},
         year = 2025,
        month = nov,
       volume = {993},
       number = {1},
          eid = {L23},
        pages = {L23},
          doi = {10.3847/2041-8213/ae1232},
archivePrefix = {arXiv},
       eprint = {2509.01647},
 primaryClass = {astro-ph.EP},
       adsurl = {https://ui.adsabs.harvard.edu/abs/2025ApJ...993L..23S}
}

@ARTICLE{Rahatgaonkar2025,
       author = {{Rahatgaonkar}, Rohan and {Carvajal}, Juan Pablo and {Puzia}, Thomas H. and {Luco}, Baltasar and {Jehin}, Emmanuel and {Hutsem{\'e}kers}, Damien and {Opitom}, Cyrielle and {Manfroid}, Jean and {Aravind}, K. and {Marsset}, Micha{\"e}l and {Yang}, Bin and {Buchanan}, Laura and {Fraser}, Wesley C. and {Forbes}, John and {Bannister}, Michele and {Bodewits}, Dennis and {Bolin}, Bryce T. and {Belyakov}, Matthew and {Knight}, Matthew M. and {Snodgrass}, Colin and {Bufanda}, Erica and {Dorsey}, Rosemary and {Ferellec}, L{\'e}a and {La Forgia}, Fiorangela and {Lippi}, Manuela and {Murphy}, Brian and {Nayak}, Prasanta K. and {Vander Donckt}, Mathieu},
        title = "{Very Large Telescope Observations of Interstellar Comet 3I/ATLAS. II. From Quiescence to Glow: Dramatic Rise of Ni I Emission and Incipient CN Outgassing at Large Heliocentric Distances}",
      journal = {\apjl},
         year = 2025,
        month = dec,
       volume = {995},
       number = {1},
          eid = {L34},
        pages = {L34},
          doi = {10.3847/2041-8213/ae1cbc},
archivePrefix = {arXiv},
       eprint = {2508.18382},
 primaryClass = {astro-ph.SR},
       adsurl = {https://ui.adsabs.harvard.edu/abs/2025ApJ...995L..34R}
}

@ARTICLE{Xing2025,
       author = {{Xing}, Zexi and {Oset}, Shawn and {Noonan}, John and {Bodewits}, Dennis},
        title = "{Water Production Rates of the Interstellar Object 3I/ATLAS}",
      journal = {\apjl},
         year = 2025,
        month = oct,
       volume = {991},
       number = {2},
          eid = {L50},
        pages = {L50},
          doi = {10.3847/2041-8213/ae08ab},
archivePrefix = {arXiv},
       eprint = {2508.04675},
 primaryClass = {astro-ph.EP},
       adsurl = {https://ui.adsabs.harvard.edu/abs/2025ApJ...991L..50X}
}

@ARTICLE{Puzia2025,
       author = {{Puzia}, Thomas H. and {Rahatgaonkar}, Rohan and {Carvajal}, Juan Pablo and {Nayak}, Prasanta K. and {Luco}, Baltasar},
        title = "{Spectral Characteristics of Interstellar Object 3I/ATLAS from SOAR Observations}",
      journal = {\apjl},
         year = 2025,
        month = sep,
       volume = {990},
       number = {1},
          eid = {L27},
        pages = {L27},
          doi = {10.3847/2041-8213/adfa0b},
archivePrefix = {arXiv},
       eprint = {2508.02777},
 primaryClass = {astro-ph.EP},
       adsurl = {https://ui.adsabs.harvard.edu/abs/2025ApJ...990L..27P}
}

@ARTICLE{Santana-Ros2025,
       author = {{Santana-Ros}, T. and {Ivanova}, O. and {Mykhailova}, S. and {Erasmus}, N. and {Kami{\'n}ski}, K. and {Oszkiewicz}, D. and {Kwiatkowski}, T. and {Hus{\'a}rik}, M. and {Ngwane}, T.~S. and {Penttil{\"a}}, A.},
        title = "{Temporal evolution of the third interstellar comet 3I/ATLAS: Spin, color, spectra, and dust activity}",
      journal = {\aap},
         year = 2025,
        month = oct,
       volume = {702},
          eid = {L3},
        pages = {L3},
          doi = {10.1051/0004-6361/202556717},
archivePrefix = {arXiv},
       eprint = {2508.00808},
 primaryClass = {astro-ph.EP},
       adsurl = {https://ui.adsabs.harvard.edu/abs/2025A&A...702L...3S}
}

@ARTICLE{Feinstein2025,
       author = {{Feinstein}, Adina D. and {Noonan}, John W. and {Seligman}, Darryl Z.},
        title = "{Precovery Observations of 3I/ATLAS from TESS Suggest Possible Distant Activity}",
      journal = {\apjl},
         year = 2025,
        month = sep,
       volume = {991},
       number = {1},
          eid = {L2},
        pages = {L2},
          doi = {10.3847/2041-8213/adfd4d},
archivePrefix = {arXiv},
       eprint = {2507.21967},
 primaryClass = {astro-ph.EP},
       adsurl = {https://ui.adsabs.harvard.edu/abs/2025ApJ...991L...2F}
}

@ARTICLE{Yang2025,
       author = {{Yang}, Bin and {Meech}, Karen J. and {Connelley}, Michael and {Zhao}, Ruining and {Keane}, Jacqueline V.},
        title = "{Spectroscopic Characterization of Interstellar Object 3I/ATLAS: Water Ice in the Coma}",
      journal = {\apjl},
         year = 2025,
        month = oct,
       volume = {992},
       number = {1},
          eid = {L9},
        pages = {L9},
          doi = {10.3847/2041-8213/ae08a7},
archivePrefix = {arXiv},
       eprint = {2507.14916},
 primaryClass = {astro-ph.EP},
       adsurl = {https://ui.adsabs.harvard.edu/abs/2025ApJ...992L...9Y}
}

@ARTICLE{Chandler2025,
       author = {{Chandler}, Colin Orion and {Bernardinelli}, Pedro H. and {Juri{\'c}}, Mario and {Singh}, Devanshi and {Hsieh}, Henry H. and {Sullivan}, Ian and {Jones}, R. Lynne and {Kurlander}, Jacob A. and {Vavilov}, Dmitrii and {Eggl}, Siegfried and {Holman}, Matthew and {Spoto}, Federica and {Schwamb}, Megan E. and {MacArthur}, Lauren A. and {Makadia}, Rahil and {Micheli}, Marco and {Heinze}, Aren and {Christensen}, Eric J. and {Beebe}, Wilson and {Roodman}, Aaron and {Lim}, Kian-Tat and {Jenness}, Tim and {Bosch}, James and {Smart}, Brianna and {Bellm}, Eric and {MacBride}, Sean and {Rawls}, Meredith L. and {Greenstreet}, Sarah and {Slater}, Colin and {Ivezi{\'c}}, {\v{Z}}eljko and {Blum}, Bob and {Connolly}, Andrew and {Daues}, Gregory and {Gower}, Michelle and {Bryce Kalmbach}, J. and {Bannister}, Michele T. and {Dones}, Luke and {Dorsey}, Rosemary C. and {Farnocchia}, Davide and {Fraser}, Wesley C. and {Forbes}, John C. and {Fuentes}, Cesar and {Holt}, Carrie E. and {Inno}, Laura and {Jones}, Geraint H. and {Knight}, Matthew M. and {Lintott}, Chris J. and {Lister}, Tim and {Lupton}, Robert and {Mendoza Magbanua}, Mark Jesus and {Malhotra}, Renu and {Mueller}, Beatrice E.~A. and {Murtagh}, Joseph and {Pandey}, Nitya and {Reach}, William T. and {Samarasinha}, Nalin H. and {Seligman}, Darryl Z. and {Snodgrass}, Colin and {Solontoi}, Michael and {Szab{\'o}}, Gyula M. and {Vere{\v{s}}}, Peter and {White}, Ellie and {Womack}, Maria and {Young}, Leslie A. and {Allbery}, Russ and {Anand}, Shreya and {Armellin}, Roberto and {Aubourg}, {\'E}ric and {Avdellidou}, Chrysa and {Azfar}, Farrukh and {Bauer}, James and {Bechtol}, Keith and {Becker{\'c}}, vValerie R and {Belyakov}, Matthew and {Benecchi}, Susan D. and {Bertini}, Ivano and {Bodewits}, Dennis and {Boeshaar}, Patricia and {Bolin}, Bryce T. and {Bose}, vMaitrayee and {Buchanan}, Laura E. and {Boucaud}, Alexandre and {Boufleur}, Rodrigo C. and {Boutigny}, Dominique and {Bradshaw}, Andrew and {Braga-Ribas}, Felipe and {Bregeon}, Johan and {Calabrese}, Daniel and {Camargo}, J.~I.~B. and {Caplar}, Neven and {Carlin}, Jeffrey L. and {Carry}, Benoit and {Carvajal}, Juan Pablo and {Ceballo}, Ross and {Chiang}, Hsin-Fang and {Choi}, Yumi and {Toribio San Cipriano}, Laura and {Combet}, C{\'e}line and {da Costa}, Luiz and {Cowan}, Preeti and {Crenshaw}, John Franklin and {Croft}, Steve and {{\'C}uk}, Matija and {Daly}, Philip N. and {Daruich}, Felipe and {Daubard}, Guillaume and {Davenport}, James R.~A. and {Daylan}, Tansu and {Delgado}, Jennifer and {Devillepoix}, Hadrien A.~R. and {Doherty}, Peter E. and {Donaldson}, Abbie and {Drass}, Holger and {Deppe}, Stephanie JH and {Dubois-Felsmann}, Gregory P. and {Ferguson}, Peter S. and {Economou}, Frossie and {Eduardo}, Marielle R. and {Sotuela Elorriaga}, Ioana and {Englert}, Anthony and {Edward-Karavakis} and {Fanning}, Kevin and {Filippo}, D'Ammando and {Frissell}, Maxwell K. and {Fedorets}, Grigori and {Fernandes}, Maryann Benny and {Fert{\'e}}, {\v{A}}gn{\`e}s and {Freytag}, Mark L and {Fulle}, Marco and {Gates{\'c}}, vJohn and {Gerdes}, David W. and {Gibbs}, Alex R. and {Gillan}, A. Fraser and {Glanzman}, T. and {Guy}, Leanne P. and {Hammergren}, Mark and {Hanushevsky}, Andrew and {Hernandez}, Fabio and {Herrold}, {\v{A}}dis and {Hestroffer}, Daniel and {Hoblitt}, Joshua and {Megias Homar}, Guillem and {Hopkins}, Matthew J. and {Giordano Orsini}, Massimiliano and {Goodenow}, Iain and {Gorsuch}, Miranda R. and {Granvik}, Mikael and {Guan}, Wen and {Le Guillou}, Laurent and {Ieva}, Simone and {Ingraham}, Patrick and {Irving}, David H. and {Jacques}, {\v{S}}ebag and {Jannuzi}, Buell T. and {Jee}, M. James and {Jimenez}, David and {Ramos Gomes-J{\'u}nior}, Altair and {Juramy}, Claire and {Kahn}, Steven M. and {Kannawadi}, Arun and {Kang}, Yijung and {Kavelaars}, JJ and {Kelley}, Michael S.~P. and {Kelkar}, Kshitija and {Kelvin}, Lee S. and {Kryszczy{\'n}ska}, Agnieszka and {Kotov}, Ivan and {Koumjian}, Alec and {Kov{\'a}cs}, G{\'a}bor and {Krughoff}, K. Simon and {Kub{\'a}nek}, Petr and {Lage}, Craig and {Lange}, Travis J. and {L{\'e}get}, Pierre-Fran{\c{c}}ois and {Fisher-Levine}, Merlin and {Levine}, Benjamin and {Levine}, W. Garrett and {Li}, Zhuofu and {Liang}, Shuang and {Licandro}, Javier and {Liss{\'e}}, {\v{C}}arey and {Lust}, Nate B. and {Lyttle}, Ryan R. and {Mainetti}, Gabriele and {Mahabal}, Ashish A. and {Mahlke}, Max and {Plazas Malag{\'o}n}, Andr{\'e}s A. and {Mandelbaum}, Rachel and {Salazar Manzano}, Luis E. and {Marc}, Moniez and {Margheim}, Steven J. and {Margoti}, Giuliano and {Morales Mar{\'\i}n C.~A.}, L. and {Mar{\v{c}}eta}, Du{\v{s}}an and {Melita}, Mario D. and {Menanteau}, Felipe and {Meyers}, Joshua and {Millsc}, Dave and {Morato}, Naomi and {More}, Surhud},
        title = "{NSF-DOE Vera C. Rubin Observatory Observations of Interstellar Comet 3I/ATLAS (C/2025 N1)}",
      journal = {arXiv e-prints},
         year = 2025,
        month = jul,
          eid = {arXiv:2507.13409},
        pages = {arXiv:2507.13409},
          doi = {10.48550/arXiv.2507.13409},
archivePrefix = {arXiv},
       eprint = {2507.13409},
 primaryClass = {astro-ph.EP},
       adsurl = {https://ui.adsabs.harvard.edu/abs/2025arXiv250713409C}
}

@ARTICLE{delaFuenteMarcos2025,
       author = {{de la Fuente Marcos}, R. and {Alarcon}, M.~R. and {Licandro}, J. and {Serra-Ricart}, M. and {de Le{\'o}n}, J. and {de la Fuente Marcos}, C. and {Lombardi}, G. and {Tejero}, A. and {Cabrera-Lavers}, A. and {Guerra Arencibia}, S. and {Ruiz Cejudo}, I.},
        title = "{Assessing interstellar comet 3I/ATLAS with the 10.4 m Gran Telescopio Canarias and the Two-meter Twin Telescope}",
      journal = {\aap},
         year = 2025,
        month = aug,
       volume = {700},
          eid = {L9},
        pages = {L9},
          doi = {10.1051/0004-6361/202556439},
archivePrefix = {arXiv},
       eprint = {2507.12922},
 primaryClass = {astro-ph.EP},
       adsurl = {https://ui.adsabs.harvard.edu/abs/2025A&A...700L...9D}
}

@ARTICLE{Kareta2025,
       author = {{Kareta}, Theodore and {Champagne}, Chansey and {McClure}, Lucas and {Emery}, Joshua and {Sharkey}, Benjamin N.~L. and {Bauer}, James and {Connelley}, Michael S. and {Rayner}, John and {Thomas}, Cristina A. and {Reddy}, Vishnu and {Firgard}, Megan},
        title = "{Near-discovery Observations of Interstellar Comet 3I/ATLAS with the NASA Infrared Telescope Facility}",
      journal = {\apjl},
         year = 2025,
        month = sep,
       volume = {990},
       number = {2},
          eid = {L65},
        pages = {L65},
          doi = {10.3847/2041-8213/adfbdf},
archivePrefix = {arXiv},
       eprint = {2507.12234},
 primaryClass = {astro-ph.EP},
       adsurl = {https://ui.adsabs.harvard.edu/abs/2025ApJ...990L..65K}
}

@ARTICLE{Belyakov2025,
       author = {{Belyakov}, Matthew and {Fremling}, Christoffer and {Graham}, Matthew J. and {Bolin}, Bryce T. and {Kilic}, Mukremin and {Jewett}, Gracyn and {Lisse}, Carey M. and {Ingebretsen}, Carl and {Davis}, M. Ryleigh and {Wong}, Ian},
        title = "{Palomar and Apache Point Spectrophotometry of Interstellar Comet 3I/ATLAS}",
      journal = {Research Notes of the American Astronomical Society},
         year = 2025,
        month = jul,
       volume = {9},
       number = {7},
          eid = {194},
        pages = {194},
          doi = {10.3847/2515-5172/adf059},
archivePrefix = {arXiv},
       eprint = {2507.11720},
 primaryClass = {astro-ph.EP},
       adsurl = {https://ui.adsabs.harvard.edu/abs/2025RNAAS...9..194B}
}

@ARTICLE{Alvarez-Candal2025,
       author = {{Alvarez-Candal}, A. and {Rizos}, J.~L. and {Lara}, L.~M. and {Santos-Sanz}, P. and {Gutierrez}, P.~J. and {Ortiz}, J.~L. and {Morales}, N.},
        title = "{X-SHOOTER spectrum of comet 3I/ATLAS: Insights into a distant interstellar visitor}",
      journal = {\aap},
         year = 2025,
        month = aug,
       volume = {700},
          eid = {L10},
        pages = {L10},
          doi = {10.1051/0004-6361/202556338},
archivePrefix = {arXiv},
       eprint = {2507.07312},
 primaryClass = {astro-ph.EP},
       adsurl = {https://ui.adsabs.harvard.edu/abs/2025A&A...700L..10A}
}

@ARTICLE{Hopkins2025,
       author = {{Hopkins}, Matthew J. and {Dorsey}, Rosemary C. and {Forbes}, John C. and {Bannister}, Michele T. and {Lintott}, Chris J. and {Leicester}, Brayden},
        title = "{From a Different Star: 3I/ATLAS in the Context of the {\={O}}tautahi{\textendash}Oxford Interstellar Object Population Model}",
      journal = {\apjl},
         year = 2025,
        month = sep,
       volume = {990},
       number = {2},
          eid = {L30},
        pages = {L30},
          doi = {10.3847/2041-8213/adfbf4},
archivePrefix = {arXiv},
       eprint = {2507.05318},
 primaryClass = {astro-ph.EP},
       adsurl = {https://ui.adsabs.harvard.edu/abs/2025ApJ...990L..30H}
}

@ARTICLE{Opitom2025,
       author = {{Opitom}, Cyrielle and {Snodgrass}, Colin and {Jehin}, Emmanuel and {Bannister}, Michele T. and {Bufanda}, Erica and {Deam}, Sophie E. and {Dorsey}, Rosemary C. and {Ferrais}, Marin and {Hmiddouch}, Said and {Knight}, Matthew M. and {Kokotanekova}, Rosita and {Leicester}, Brayden and {Marsset}, Micha{\"e}l and {Murphy}, Brian and {Okoth}, Vincent and {Ridden-Harper}, Ryan and {Vander Donckt}, Mathieu and {Ferellec}, L{\'e}a and {Hutsem{\'e}kers}, Damien and {Lippi}, Manuela and {Manfroid}, Jean and {Benkhaldoun}, Zouhair},
        title = "{Snapshot of a new interstellar comet: 3I/ATLAS has a red and featureless spectrum}",
      journal = {\mnras},
         year = 2025,
        month = nov,
       volume = {544},
       number = {1},
        pages = {L31-L36},
          doi = {10.1093/mnrasl/slaf095},
archivePrefix = {arXiv},
       eprint = {2507.05226},
 primaryClass = {astro-ph.EP},
       adsurl = {https://ui.adsabs.harvard.edu/abs/2025MNRAS.544L..31O}
}

@ARTICLE{Frincke2026,
       author = {{Frincke}, Tessa T. and {Yaginuma}, Atsuhiro and {Noonan}, John W. and {Hsieh}, Henry H. and {Seligman}, Darryl Z. and {Holt}, Carrie E. and {Strader}, Jay and {Do}, Thomas and {Craig}, Peter and {Molina}, Isabella},
        title = "{Near-discovery SOAR photometry of the third interstellar object: 3I/ATLAS}",
      journal = {\mnras},
         year = 2026,
        month = jan,
       volume = {545},
       number = {1},
          eid = {staf1994},
        pages = {staf1994},
          doi = {10.1093/mnras/staf1994},
archivePrefix = {arXiv},
       eprint = {2509.02813},
 primaryClass = {astro-ph.EP},
       adsurl = {https://ui.adsabs.harvard.edu/abs/2026MNRAS.545f1994F}
}

@ARTICLE{Lisse2025b,
       author = {{Lisse}, C.~M. and {Bach}, Y.~P. and {Bryan}, S. and {Crill}, B.~P. and {Cukierman}, A. and {Dor{\'e}}, O. and {Fabinsky}, B. and {Faisst}, A. and {Korngut}, P.~M. and {Melnick}, G. and {Rustamkulov}, Z. and {Tolls}, V. and {Werner}, M. and {Sitko}, M.~L. and {Champagne}, C. and {Connelley}, M. and {Emery}, J.~P. and {Fernandez}, Y.~R. and {Yang}, B. and {the SPHEREx Science Team}},
        title = "{SPHEREx Discovery of Strong Water Ice Absorption and an Extended Carbon Dioxide Coma in 3I/ATLAS}",
      journal = {Research Notes of the American Astronomical Society},
         year = 2025,
        month = sep,
       volume = {9},
       number = {9},
          eid = {242},
        pages = {242},
          doi = {10.3847/2515-5172/ae0293},
archivePrefix = {arXiv},
       eprint = {2508.15469},
 primaryClass = {astro-ph.EP},
       adsurl = {https://ui.adsabs.harvard.edu/abs/2025RNAAS...9..242L}
}

@ARTICLE{Fraser-Gillan2026,
       author = {{Gillan}, A. Fraser and {Wyrzykowski}, {\L}ukasz and {Miko{\l}ajczyk}, Przemys{\l}aw J. and {Kotysz}, Krzysztof and {Bufanda}, Erica and {Chandler}, Colin O. and {Fi{\textcommabelow s}ek}, S{\"u}leyman and {Hsieh}, Henry H. and {Kelley}, Michael S.~P. and {Pessi}, Priscila J. and {Robinson}, James E. and {Ali{\textcommabelow s}}, Sinan and {Bykowski}, Wie{\'n}czys{\l}aw and {Cannon}, Richard E. and {Dominik}, Martin and {Handzlik}, Barbara and {{\.I}{\c{c}}en}, Mehmet and {Kurowski}, Sebastian and {Cem Kutluay}, Ahmet and {Majumdar}, Joysankar and {Nehir}, {\c{C}}a{\u{g}}layan and {O'Neill}, David and {{\"O}tken}, Sibel and {Pu}, Kangming and {{\c{S}}im{\textcommabelow s}ir}, {\"O}zlem and {Snodgrass}, Colin and {Tu{\u{g}}rul Tezcan}, Cihan and {Tezcan}, Fatma and {Wicker}, Mauritz and {Yelkenci}, Fuat Korhan and {{\.Z}ejmo}, Micha{\l} and {Ackley}, Kendall and {Andersen}, M. and {{\'A}valos-Vega}, C. and {Belkin}, Sergey and {Bozza}, V. and {Breton}, Rene P. and {Burgdorf}, M.~J. and {Casares}, Jorge and {Dhillon}, Vik and {Donaldson}, A. and {Dyer}, Martin J. and {Figuera Jaimes}, R. and {Galloway}, Duncan K. and {Hinse}, T.~C. and {Hundertmark}, M. and {Khalouei}, E. and {Killestein}, Thomas and {Kotak}, Rubina and {Kumar}, Amit and {Frey Liu}, Feng-Yuan and {Longa-Pe{\~n}a}, P. and {Lyman}, Joe and {Mancini}, Luigi and {Moharana}, A. and {Molina}, V. and {Noysena}, Kanthanakorn and {Nuttall}, Laura Kate and {O'Brien}, Paul and {Okoth}, V. and {Opitom}, C. and {Pollacco}, Don and {Rabus}, M. and {Ramsay}, Gavin and {Sajadian}, S. and {Salinas San Martin}, A. and {Skottfelt}, J. and {Southworth}, J. and {Steeghs}, Danny and {Tregloan-Reed}, J. and {Ulaczyk}, Krzysztof and {Vieliute}, R.},
        title = "{Time-Domain Photometry and Activity Evolution of Interstellar Comet 3I/ATLAS with BHTOM}",
      journal = {arXiv e-prints},
         year = 2026,
        month = mar,
          eid = {arXiv:2603.01383},
        pages = {arXiv:2603.01383},
          doi = {10.48550/arXiv.2603.01383},
archivePrefix = {arXiv},
       eprint = {2603.01383},
 primaryClass = {astro-ph.EP},
       adsurl = {https://ui.adsabs.harvard.edu/abs/2026arXiv260301383G}
}

@ARTICLE{Cordiner2026,
       author = {{Cordiner}, Martin and {Roth}, Nathan X. and {Micheli}, Marco and {Villanueva}, Geronimo and {Farnocchia}, Davide and {Charnley}, Steven and {Biver}, Nicolas and {Bockelee-Morvan}, Dominique and {Bodewits}, Dennis and {Chandler}, Colin Orion and {Crovisier}, Jacques and {Drozdovskaya}, Maria N. and {Furuya}, Kenji and {Kelley}, Michael S.~P. and {Milam}, Stefanie and {Noonan}, John W. and {Opitom}, Cyrielle and {Schwamb}, Megan E. and {Thomas}, Cristina A.},
        title = "{Isotopic Evidence for a Cold and Distant Origin of the Interstellar Object 3I/ATLAS}",
      journal = {arXiv e-prints},
         year = 2026,
        month = mar,
          eid = {arXiv:2603.06911},
        pages = {arXiv:2603.06911},
          doi = {10.48550/arXiv.2603.06911},
archivePrefix = {arXiv},
       eprint = {2603.06911},
 primaryClass = {astro-ph.EP},
       adsurl = {https://ui.adsabs.harvard.edu/abs/2026arXiv260306911C}
}

@article{Jewitt2017,
doi = {10.3847/2041-8213/aa9b2f},
url = {https://doi.org/10.3847/2041-8213/aa9b2f},
year = {2017},
month = {nov},
publisher = {The American Astronomical Society},
volume = {850},
number = {2},
pages = {L36},
author = {Jewitt, David and Luu, Jane and Rajagopal, Jayadev and Kotulla, Ralf and Ridgway, Susan and Liu, Wilson and Augusteijn, Thomas},
title = {Interstellar Interloper 1I/2017 U1: Observations from the NOT and WIYN Telescopes},
journal = {The Astrophysical Journal Letters}
}

@ARTICLE{Zhao2026,
       author = {{Zhao}, Ruining and {Zhang}, Xiliang and {Yang}, Bin and {Fan}, Xiangyu and {Wang}, Shu and {Huang}, Yang and {Liu}, Jifeng},
        title = "{Post-perihelion Coma Composition of the Interstellar Comet 3I/ATLAS from Optical Spectroscopy}",
      journal = {arXiv e-prints},
         year = 2026,
        month = mar,
          eid = {arXiv:2603.07718},
        pages = {arXiv:2603.07718},
          doi = {10.48550/arXiv.2603.07718},
archivePrefix = {arXiv},
       eprint = {2603.07718},
 primaryClass = {astro-ph.EP},
       adsurl = {https://ui.adsabs.harvard.edu/abs/2026arXiv260307718Z}
}

@ARTICLE{Roth2026,
       author = {{Roth}, Nathan X. and {Cordiner}, Martin and {Milam}, Stefanie and {Villanueva}, Geronimo and {Charnley}, Steven and {Biver}, Nicolas and {Bockelee-Morvan}, Dominique and {Bodewits}, Dennis and {Crovisier}, Jacques and {Drozdovskaya}, Maria N. and {Farnocchia}, Davide and {Furuya}, Kenji and {Kelley}, Michael S.~P. and {Micheli}, Marco and {Noonan}, John W. and {Opitom}, Cyrielle and {Schwamb}, Megan E. and {Thomas}, Cristina A.},
        title = "{An Enriched Methane D/H Ratio in the Interstellar Object 3I/ATLAS}",
      journal = {arXiv e-prints},
         year = 2026,
        month = mar,
          eid = {arXiv:2603.20445},
        pages = {arXiv:2603.20445},
          doi = {10.48550/arXiv.2603.20445},
archivePrefix = {arXiv},
       eprint = {2603.20445},
 primaryClass = {astro-ph.EP},
       adsurl = {https://ui.adsabs.harvard.edu/abs/2026arXiv260320445R}
}

@ARTICLE{Biver2026,
       author = {{Biver}, N. and {Bockel{\'e}e-Morvan}, D. and {Moreno}, R. and {Crovisier}, J. and {Paubert}, G. and {Zakharov}, V. and {Boissier}, J. and {Cordiner}, M.~A. and {Roth}, N.~X.},
        title = "{Perihelion observations of interstellar comet 3I/ATLAS with the IRAM 30-m telescope}",
      journal = {arXiv e-prints},
         year = 2026,
        month = mar,
          eid = {arXiv:2603.23240},
        pages = {arXiv:2603.23240},
          doi = {10.48550/arXiv.2603.23240},
archivePrefix = {arXiv},
       eprint = {2603.23240},
 primaryClass = {astro-ph.EP},
       adsurl = {https://ui.adsabs.harvard.edu/abs/2026arXiv260323240B}
}

@ARTICLE{Shinnaka2026,
       author = {{Shinnaka}, Yoshiharu and {Tsujimoto}, Ko and {Kawakita}, Hideyo and {Kobayashi}, Hitomi and {Watanabe}, Jun-ichi and {Ootsubo}, Takafumi},
        title = "{A post-perihelion constraint on the CO$_{2}$/H$_{2}$O ratio of interstellar comet 3I/ATLAS from [O I] forbidden lines}",
      journal = {arXiv e-prints},
         year = 2026,
        month = mar,
          eid = {arXiv:2603.25002},
        pages = {arXiv:2603.25002},
          doi = {10.48550/arXiv.2603.25002},
archivePrefix = {arXiv},
       eprint = {2603.25002},
 primaryClass = {astro-ph.GA},
       adsurl = {https://ui.adsabs.harvard.edu/abs/2026arXiv260325002S}
}

@ARTICLE{Hutsemekers2026b,
       author = {{Hutsem{\'e}kers}, Damien and {Manfroid}, Jean and {Opitom}, Cyrielle and {Jehin}, Emmanu{\"e}l and {Krishnakumar}, Aravind and {Massa Fernandes}, Fernando and {Bannister}, Michele and {Bodewits}, Dennis and {Dorsey}, Rosemary and {La Forgia}, Fiorangela and {Murphy}, Brian},
        title = "{Origin and evolution of NiI and FeI in the coma of the interstellar comet 3I/ATLAS throughout its trajectory}",
      journal = {arXiv e-prints},
         year = 2026,
        month = may,
          eid = {arXiv:2605.07652},
        pages = {arXiv:2605.07652},
archivePrefix = {arXiv},
       eprint = {2605.07652},
 primaryClass = {astro-ph.EP},
       adsurl = {https://ui.adsabs.harvard.edu/abs/2026arXiv260507652H}
}

@ARTICLE{Williams17,
       author = {{Williams}, G.~V. and {Sato}, H. and {Sarneczky}, K. and {Wainscoat}, R. and {Woodworth}, D. and {Meech}, K.},
        title = "{Minor Planets 2017 SN\_33 and 2017 U1}",
      journal = {Central Bureau Electronic Telegrams},
         year = 2017,
        month = oct,
       volume = {4450},
        pages = {1},
       adsurl = {https://ui.adsabs.harvard.edu/abs/2017CBET.4450....1W}
}

@ARTICLE{borisov_2I_cbet,
       author = {{Borisov}, G. and {Durig}, D. T. and {Sato}, H. and {Birtwhistle}, P. and {Chen}, T. and {Green}, D. W. E. and {Bacci}, P. and {Maestripieri}, M. and {Nakano}, S.},
        title = "{Comet C/2019 Q4 (Borisov)}",
      journal = {Central Bureau Electronic Telegrams},
         year = "2019",
        month = sep,
       volume = {4666},
        pages = {1}
}

@ARTICLE{Hui2020,
       author = {{Hui}, Man-To and {Ye}, Quan-Zhi and {F{\"o}hring}, Dora and {Hung}, Denise and {Tholen}, David J.},
        title = "{Physical Characterization of Interstellar Comet 2I/2019 Q4 (Borisov)}",
      journal = {\aj},
         year = 2020,
        month = aug,
       volume = {160},
       number = {2},
          eid = {92},
        pages = {92},
          doi = {10.3847/1538-3881/ab9df8},
       adsurl = {https://ui.adsabs.harvard.edu/abs/2020AJ....160...92H}
}

@ARTICLE{Guzik2020,
       author = {{Guzik}, Piotr and {Drahus}, Micha{\l} and {Rusek}, Krzysztof and
         {Waniak}, Wac{\l}aw and {Cannizzaro}, Giacomo and
         {Pastor-Marazuela}, In{\'e}s},
        title = "{Initial characterization of interstellar comet 2I/Borisov}",
      journal = {Nature Astronomy},
         year = "2020",
        month = "Jan",
       volume = {4},
        pages = {53-57},
          doi = {10.1038/s41550-019-0931-8},
archivePrefix = {arXiv},
       eprint = {1909.05851},
 primaryClass = {astro-ph.EP},
       adsurl = {https://ui.adsabs.harvard.edu/abs/2020NatAs...4...53G}
}

@ARTICLE{Jewitt2019b,
       author = {{Jewitt}, David and {Luu}, Jane},
        title = "{Initial Characterization of Interstellar Comet 2I/2019 Q4 (Borisov)}",
      journal = {\apjl},
         year = "2019",
        month = "Dec",
       volume = {886},
       number = {2},
          eid = {L29},
        pages = {L29},
          doi = {10.3847/2041-8213/ab530b},
archivePrefix = {arXiv},
       eprint = {1910.02547},
 primaryClass = {astro-ph.EP},
       adsurl = {https://ui.adsabs.harvard.edu/abs/2019ApJ...886L..29J}
}

@ARTICLE{Ye2017,
       author = {{Ye}, Quan-Zhi and {Zhang}, Qicheng and {Kelley}, Michael S.~P. and {Brown}, Peter G.},
        title = "{1I/2017 U1 ({\textquoteleft}Oumuamua) is Hot: Imaging, Spectroscopy, and Search of Meteor Activity}",
      journal = {\apjl},
         year = 2017,
        month = dec,
       volume = {851},
       number = {1},
          eid = {L5},
        pages = {L5},
          doi = {10.3847/2041-8213/aa9a34},
archivePrefix = {arXiv},
       eprint = {1711.02320},
 primaryClass = {astro-ph.EP},
       adsurl = {https://ui.adsabs.harvard.edu/abs/2017ApJ...851L...5Y}
}

@ARTICLE{McKay2020,
       author = {{McKay}, Adam J. and {Cochran}, Anita L. and {Dello Russo}, Neil and {DiSanti}, Michael A.},
        title = "{Detection of a Water Tracer in Interstellar Comet 2I/Borisov}",
      journal = {\apjl},
         year = 2020,
        month = jan,
       volume = {889},
       number = {1},
          eid = {L10},
        pages = {L10},
          doi = {10.3847/2041-8213/ab64ed},
archivePrefix = {arXiv},
       eprint = {1910.12785},
 primaryClass = {astro-ph.EP},
       adsurl = {https://ui.adsabs.harvard.edu/abs/2020ApJ...889L..10M}
}

@ARTICLE{Fitzsimmons2019,
       author = {{Fitzsimmons}, Alan and {Hainaut}, Olivier and {Meech}, Karen J. and {Jehin}, Emmanuel and {Moulane}, Youssef and {Opitom}, Cyrielle and {Yang}, Bin and {Keane}, Jacqueline V. and {Kleyna}, Jan T. and {Micheli}, Marco and {Snodgrass}, Colin},
        title = "{Detection of CN Gas in Interstellar Object 2I/Borisov}",
      journal = {\apjl},
         year = 2019,
        month = nov,
       volume = {885},
       number = {1},
          eid = {L9},
        pages = {L9},
          doi = {10.3847/2041-8213/ab49fc},
archivePrefix = {arXiv},
       eprint = {1909.12144},
 primaryClass = {astro-ph.EP},
       adsurl = {https://ui.adsabs.harvard.edu/abs/2019ApJ...885L...9F}
}

@ARTICLE{Kim2020,
       author = {{Kim}, Yoonyoung and {Jewitt}, David and {Mutchler}, Max and {Agarwal}, Jessica and {Hui}, Man-To and {Weaver}, Harold},
        title = "{Coma Anisotropy and the Rotation Pole of Interstellar Comet 2I/Borisov}",
      journal = {\apjl},
         year = 2020,
        month = jun,
       volume = {895},
       number = {2},
          eid = {L34},
        pages = {L34},
          doi = {10.3847/2041-8213/ab9228},
archivePrefix = {arXiv},
       eprint = {2005.02468},
 primaryClass = {astro-ph.EP},
       adsurl = {https://ui.adsabs.harvard.edu/abs/2020ApJ...895L..34K}
}

@ARTICLE{Cremonese2020,
       author = {{Cremonese}, G. and {Fulle}, M. and {Cambianica}, P. and {Munaretto}, G. and {Capria}, M.~T. and {La Forgia}, F. and {Lazzarin}, M. and {Migliorini}, A. and {Boschin}, W. and {Milani}, G. and {Aletti}, A. and {Arlic}, G. and {Bacci}, P. and {Bacci}, R. and {Bryssinck}, E. and {Carosati}, D. and {Castellano}, D. and {Buzzi}, L. and {Di Rubbo}, S. and {Facchini}, M. and {Guido}, E. and {Kugel}, F. and {Ligustri}, R. and {Maestripieri}, M. and {Mantero}, A. and {Nicolas}, J. and {Ochner}, P. and {Perrella}, C. and {Trabatti}, R. and {Valvasori}, A.},
        title = "{Dust Environment Model of the Interstellar Comet 2I/Borisov}",
      journal = {\apjl},
         year = 2020,
        month = apr,
       volume = {893},
       number = {1},
          eid = {L12},
        pages = {L12},
          doi = {10.3847/2041-8213/ab8455},
       adsurl = {https://ui.adsabs.harvard.edu/abs/2020ApJ...893L..12C}
}

@article{Yang2021,
	Author = {Yang, Bin and Li, Aigen and Cordiner, Martin A. and Chang, Chin-Shin and Hainaut, Olivier R. and Williams, Jonathan P. and Meech, Karen J. and Keane, Jacqueline V. and Villard, Eric},
	Da = {2021/03/30},
	Doi = {10.1038/s41550-021-01336-w},
	Id = {Yang2021},
	Isbn = {2397-3366},
	Journal = {Nature Astronomy},
	Title = {Compact pebbles and the evolution of volatiles in the interstellar comet 2I/Borisov},
	Ty = {JOUR},
	Url = {https://doi.org/10.1038/s41550-021-01336-w},
	Year = {2021}}

@ARTICLE{Opitom2019,
       author = {{Opitom}, Cyrielle and {Fitzsimmons}, Alan and {Jehin}, Emmanuel and
         {Moulane}, Youssef and {Hainaut}, Olivier and {Meech}, Karen J. and
         {Yang}, Bin and {Snodgrass}, Colin and {Micheli}, Marco and
         {Keane}, Jacqueline V. and {Benkhaldoun}, Zouhair and {Kleyna}, Jan T.},
        title = "{2I/Borisov: A C$_{2}$-depleted interstellar comet}",
      journal = {\aap},
         year = "2019",
        month = "Nov",
       volume = {631},
          eid = {L8},
        pages = {L8},
          doi = {10.1051/0004-6361/201936959},
       adsurl = {https://ui.adsabs.harvard.edu/abs/2019A&A...631L...8O}
}

@ARTICLE{Kareta2019,
       author = {{Kareta}, Theodore and {Andrews}, Jennifer and {Noonan}, John W. and {Harris}, Walter M. and {Smith}, Nathan and {O'Brien}, Patrick and {Sharkey}, Benjamin N.~L. and {Reddy}, Vishnu and {Springmann}, Alessondra and {Lejoly}, Cassandra and {Volk}, Kathryn and {Conrad}, Albert and {Veillet}, Christian},
        title = "{Carbon Chain Depletion of 2I/Borisov}",
      journal = {\apjl},
         year = 2020,
        month = feb,
       volume = {889},
       number = {2},
          eid = {L38},
        pages = {L38},
          doi = {10.3847/2041-8213/ab6a08},
archivePrefix = {arXiv},
       eprint = {1910.03222},
 primaryClass = {astro-ph.EP},
       adsurl = {https://ui.adsabs.harvard.edu/abs/2020ApJ...889L..38K}
}

@ARTICLE{Lin2020,
       author = {{Lin}, Hsing Wen and {Lee}, Chien-Hsiu and {Gerdes}, D.~W. and {Adams}, Fred C. and {Becker}, Juliette and {Napier}, Kevin and {Markwardt}, Larissa},
        title = "{Detection of Diatomic Carbon in 2I/Borisov}",
      journal = {\apjl},
         year = 2020,
        month = feb,
       volume = {889},
       number = {2},
          eid = {L30},
        pages = {L30},
          doi = {10.3847/2041-8213/ab6bd9},
archivePrefix = {arXiv},
       eprint = {1912.06161},
 primaryClass = {astro-ph.EP},
       adsurl = {https://ui.adsabs.harvard.edu/abs/2020ApJ...889L..30L}
}

@ARTICLE{Xing2020,
       author = {{Xing}, Zexi and {Bodewits}, Dennis and {Noonan}, John and {Bannister}, Michele T.},
        title = "{Water Production Rates and Activity of Interstellar Comet 2I/Borisov}",
      journal = {\apjl},
         year = 2020,
        month = apr,
       volume = {893},
       number = {2},
          eid = {L48},
        pages = {L48},
          doi = {10.3847/2041-8213/ab86be},
archivePrefix = {arXiv},
       eprint = {2001.04865},
 primaryClass = {astro-ph.EP},
       adsurl = {https://ui.adsabs.harvard.edu/abs/2020ApJ...893L..48X}
}

@ARTICLE{Aravind2021,
       author = {{Aravind}, K. and {Ganesh}, Shashikiran and {Venkataramani}, Kumar and {Sahu}, Devendra and {Angchuk}, Dorje and {Sivarani}, Thirupathi and {Unni}, Athira},
        title = "{Activity of the first interstellar comet 2I/Borisov around perihelion: results from Indian observatories}",
      journal = {\mnras},
         year = 2021,
        month = apr,
       volume = {502},
       number = {3},
        pages = {3491-3499},
          doi = {10.1093/mnras/stab084},
archivePrefix = {arXiv},
       eprint = {2101.02752},
 primaryClass = {astro-ph.EP},
       adsurl = {https://ui.adsabs.harvard.edu/abs/2021MNRAS.502.3491A}
}

@ARTICLE{Bagnulo2021,
       author = {{Bagnulo}, S. and {Cellino}, A. and {Kolokolova}, L. and {Ne{\v{z}}i{\v{c}}}, R. and {Santana-Ros}, T. and {Borisov}, G. and {Christou}, A.~A. and {Bendjoya}, Ph. and {Devog{\`e}le}, M.},
        title = "{Unusual polarimetric properties for interstellar comet 2I/Borisov}",
      journal = {Nature Communications},
         year = 2021,
        month = jan,
       volume = {12},
          eid = {1797},
        pages = {1797},
          doi = {10.1038/s41467-021-22000-x},
       adsurl = {https://ui.adsabs.harvard.edu/abs/2021NatCo..12.1797B}
}

@ARTICLE{Deam2025,
       author = {{Deam}, Sophie E. and {Bannister}, Michele T. and {Opitom}, Cyrielle and {Knight}, Matthew M. and {Ridden-Harper}, Ryan and {Seligman}, Darryl Z. and {Fitzsimmons}, Alan and {Guilbert-Lepoutre}, Aur{\'e}lie and {Jehin}, Emmanuel and {Jorda}, Laurent and {Marsset}, Michael and {Moulane}, Youssef and {Rousselot}, Philippe and {Vernazza}, Pierre and {Yang}, Bin},
        title = "{A portrait throughout perihelion of the NH$_2$-rich interstellar comet 2I/Borisov}",
      journal = {arXiv e-prints},
         year = 2025,
        month = jul,
          eid = {arXiv:2507.05051},
        pages = {arXiv:2507.05051},
          doi = {10.48550/arXiv.2507.05051},
archivePrefix = {arXiv},
       eprint = {2507.05051},
 primaryClass = {astro-ph.EP},
       adsurl = {https://ui.adsabs.harvard.edu/abs/2025arXiv250705051D}
}

@ARTICLE{Bodewits2020,
       author = {{Bodewits}, D. and {Noonan}, J.~W. and {Feldman}, P.~D. and {Bannister}, M.~T. and {Farnocchia}, D. and {Harris}, W.~M. and {Li}, J. -Y. and {Mandt}, K.~E. and {Parker}, J. Wm. and {Xing}, Z. -X.},
        title = "{The carbon monoxide-rich interstellar comet 2I/Borisov}",
      journal = {Nature Astronomy},
         year = 2020,
        month = apr,
       volume = {4},
        pages = {867-871},
          doi = {10.1038/s41550-020-1095-2},
archivePrefix = {arXiv},
       eprint = {2004.08972},
 primaryClass = {astro-ph.EP},
       adsurl = {https://ui.adsabs.harvard.edu/abs/2020NatAs...4..867B}
}

@ARTICLE{Cordiner2020,
       author = {{Cordiner}, M.~A. and {Milam}, S.~N. and {Biver}, N. and {Bockel{\'e}e-Morvan}, D. and {Roth}, N.~X. and {Bergin}, E.~A. and {Jehin}, E. and {Remijan}, A.~J. and {Charnley}, S.~B. and {Mumma}, M.~J. and {Boissier}, J. and {Crovisier}, J. and {Paganini}, L. and {Kuan}, Y. -J. and {Lis}, D.~C.},
        title = "{Unusually high CO abundance of the first active interstellar comet}",
      journal = {Nature Astronomy},
         year = 2020,
        month = apr,
       volume = {4},
        pages = {861-866},
          doi = {10.1038/s41550-020-1087-2},
archivePrefix = {arXiv},
       eprint = {2004.09586},
 primaryClass = {astro-ph.EP},
       adsurl = {https://ui.adsabs.harvard.edu/abs/2020NatAs...4..861C}
}

@ARTICLE{Bannister2020,
       author = {{Bannister}, Michele T. and {Opitom}, Cyrielle and {Fitzsimmons}, Alan and
         {Moulane}, Youssef and {Jehin}, Emmanuel and {Seligman}, Darryl and
         {Rousselot}, Philippe and {Knight}, Matthew M. and {Marsset}, Michael and
         {Schwamb}, Megan E. and {Guilbert-Lepoutre}, Aur{\'e}lie and
         {Jorda}, Laurent and {Vernazza}, Pierre and {Benkhaldoun}, Zouhair},
        title = "{Interstellar comet 2I/Borisov as seen by MUSE: C$_2$, NH$_2$ and red CN detections}",
      journal = {arXiv e-prints},
         year = "2020",
        month = "Jan",
          eid = {arXiv:2001.11605},
        pages = {arXiv:2001.11605},
archivePrefix = {arXiv},
       eprint = {2001.11605},
 primaryClass = {astro-ph.EP},
       adsurl = {https://ui.adsabs.harvard.edu/abs/2020arXiv200111605B}
}

@ARTICLE{Manfroid2021,
       author = {{Manfroid}, J. and {Hutsem{\'e}kers}, D. and {Jehin}, E.},
        title = "{Iron and nickel atoms in cometary atmospheres even far from the Sun}",
      journal = {\nat},
         year = 2021,
        month = may,
       volume = {593},
       number = {7859},
        pages = {372-374},
          doi = {10.1038/s41586-021-03435-0},
       adsurl = {https://ui.adsabs.harvard.edu/abs/2021Natur.593..372M}
}

@ARTICLE{Kushwaha2025,
       author = {{Kushwaha}, Rahul K. and {Gudipati}, Murthy S. and {Henderson}, Bryana L.},
        title = "{Cryogenic Differential Calorimetry: Exothermicity of Amorphous-to-crystalline Phase Transitions (ACPT) in Astrophysical and Cometary Ice Analogs}",
      journal = {\apj},
         year = 2025,
        month = jul,
       volume = {987},
       number = {2},
          eid = {190},
        pages = {190},
          doi = {10.3847/1538-4357/addc68},
       adsurl = {https://ui.adsabs.harvard.edu/abs/2025ApJ...987..190K}
}

@ARTICLE{Helbert2005,
       author = {{Helbert}, J. and {Rauer}, H. and {Boice}, D.~C. and {Huebner}, W.~F.},
        title = "{The chemistry of C$_{2}$ and C$_{3}$ in the coma of Comet C/1995 O1 (Hale-Bopp) at heliocentric distances r$_{h}$ {\ensuremath{\geq}} 2.9 AU}",
      journal = {\aap},
         year = 2005,
        month = nov,
       volume = {442},
       number = {3},
        pages = {1107-1120},
          doi = {10.1051/0004-6361:20041571},
       adsurl = {https://ui.adsabs.harvard.edu/abs/2005A&A...442.1107H}
}

@ARTICLE{2020SciPy-NMeth,
  author  = {Virtanen, Pauli and Gommers, Ralf and Oliphant, Travis E. and
            Haberland, Matt and Reddy, Tyler and Cournapeau, David and
            Burovski, Evgeni and Peterson, Pearu and Weckesser, Warren and
            Bright, Jonathan and {van der Walt}, St{\'e}fan J. and
            Brett, Matthew and Wilson, Joshua and Millman, K. Jarrod and
            Mayorov, Nikolay and Nelson, Andrew R. J. and Jones, Eric and
            Kern, Robert and Larson, Eric and Carey, C J and
            Polat, {\.I}lhan and Feng, Yu and Moore, Eric W. and
            {VanderPlas}, Jake and Laxalde, Denis and Perktold, Josef and
            Cimrman, Robert and Henriksen, Ian and Quintero, E. A. and
            Harris, Charles R. and Archibald, Anne M. and
            Ribeiro, Ant{\^o}nio H. and Pedregosa, Fabian and
            {van Mulbregt}, Paul and {SciPy 1.0 Contributors}},
  title   = {{{SciPy} 1.0: Fundamental Algorithms for Scientific
            Computing in Python}},
  journal = {Nature Methods},
  year    = {2020},
  volume  = {17},
  pages   = {261--272},
  adsurl  = {https://rdcu.be/b08Wh},
  doi     = {10.1038/s41592-019-0686-2},
}

@Article{Hunter:2007,
  Author    = {Hunter, J. D.},
  Title     = {Matplotlib: A 2D graphics environment},
  Journal   = {Computing in Science \& Engineering},
  Volume    = {9},
  Number    = {3},
  Pages     = {90--95},
  publisher = {IEEE COMPUTER SOC},
  doi       = {10.1109/MCSE.2007.55},
  year      = 2007
}

@article{LinstromMallard2001,
  author = {Peter Linstrom and William Mallard},
  title = {The NIST Chemistry WebBook: A Chemical Data Resource on the Internet},
  journal = { \, },
  year = {2001},
  number = {46},
  month = {2001-09-01},
  publisher = {Journal of Chemical and Engineering Data},
  language = {en},
}

\appendix
\section{Expanded Table}
In this appendix we provide Table \ref{tab:Q-table}, which contains every reported production rate for 3I/ATLAS to date. 

\setlength{\LTcapwidth}{\textwidth}

\captionsetup[longtable]{
    labelfont=bf,
    textfont=normalfont
}
\clearpage
\onecolumn 
\begin{longtable}{lllllll}
\caption{Catalog of production rates for chemical species $Q_X$ from observations of 3I/ATLAS. Includes error associated with $Q_X$, observation dates, heliocentric distance $r_H$, facility used to obtained observation, and relevant citation. Upper limit production rates are indicated with $<$ sign and repeated entries are indicated with three dots. Note that the errors are of the same order of magnitude as the corresponding value of $Q_X$.}\\
\label{tab:Q-table}\\
\hline\hline
Volatile         & Production Rate       & Error                 & Observation Date                 & $r_H$        & Facility     & Reference                    \\ 
\endfirsthead
\endhead

\endfoot
\endlastfoot
                 & [s $^{-1}$]           & [s $^{-1}$]           &                      & [au]         &              &                              \\
\hline
H$_2$O           & $<9.10 \times 10^{26}$              & 0.00                  & 07/04/25--05/26/25   & 4.37         & X-SHOOTER    & \citet{Alvarez-Candal2025}   \\
                 & $7.40 \times 10^{26}$              & 5.00                  & 07/31/25             & 3.51         & UVOT         & \citet{Xing2025}             \\
                 & $3.20 \times 10^{26}$              & 0.20                  & 08/01/25--08/15/25   & 3.20         & SPHEREx      & \citet{Lisse2025a}           \\
                 & $2.23 \times 10^{26}$              & 0.08                  & 08/06/25             & 3.32         & JWST/NIRSpec & \citet{Cordiner2025}          \\
                 & $<1.50 \times 10^{26}$              & 0.00                  & 08/08/25--08/12/25   & 3.10--3.30   & SPHEREx      & \citet{Lisse2025b}           \\
                 & $1.36 \times 10^{27}$              & 0.50                  & 08/19/25             & 2.90         & UVOT         & \citet{Xing2025}             \\
                 & $<9.46 \times 10^{27}$              & 0.00                  & 08/26/25--09/03/25   & 2.54         & TMRT         & \citet{Li2026}               \\
\end{longtable}
\bsp	
\label{lastpage}
\end{document}